\documentclass[]{aastex631}

\usepackage{hyperref}
\usepackage{graphicx}
\usepackage{amsmath}
\usepackage{mathtools}
\usepackage{natbib}
\usepackage{amsfonts}
\usepackage{wasysym}
\usepackage{amssymb}

\accepted{July 26, 2022}

\submitjournal{ApJ}

\begin{document}

\title{DarkMix: Mixture Models for the Detection and Characterization of Dark Matter Halos}

\correspondingauthor{Llu\'is Hurtado-Gil}
\email{lluis.hurtado@edreamsodigeo.com, lluis.hurtado@uv.es}

\author[0000-0001-9674-1345]{Llu\'is Hurtado-Gil}
\affiliation{eDreams ODIGEO \\
C/ Bail\`en 67-69,\\
08009 Barcelona, Spain.}
\affiliation{Observatori Astron\`omic \\
Universitat de Val\`encia \\
C/ Catedr\`atic Jos\'e Beltr\'an 2 \\
E-46980, Paterna, Spain}

\nocollaboration{5}

\author[0000-0002-0631-7514]{Michael A. Kuhn}
\affiliation{California Institute of Technology \\
Pasadena, CA 91125, USA}

\author[0000-0003-0791-7885]{Pablo Arnalte-Mur}
\affiliation{Observatori Astron\`omic \\
Universitat de Val\`encia \\
C/ Catedr\`atic Jos\'e Beltr\'an 2 \\
E-46980, Paterna, Spain}
\affiliation{Departament d'Astronomia i Astrof\'isica \\
Universitat de Val\`encia \\
E-46100, Burjassot, Spain}

\author[0000-0002-5077-6734]{Eric D. Feigelson}
\affiliation{Department of Astronomy \& Astrophysics, \\ 
Penn State University \\
University Park, PA 16802, USA}
\affiliation{Center for Astrostatistics, \\ 
Penn State University,
University Park, PA 16802, USA}

\author[0000-0002-9937-0532]{Vicent Mart\'inez}
\affiliation{Observatori Astron\`omic \\
Universitat de Val\`encia \\
C/ Catedr\`atic Jos\'e Beltr\'an 2 \\
E-46980, Paterna, Spain}
\affiliation{Departament d'Astronomia i Astrof\'isica \\
Universitat de Val\`encia \\
E-46100, Burjassot, Spain}
\affiliation{Unidad Asociada Observatorio Astron\'omico (IFCA-UV) \\
E-46980, Paterna, Spain}

\begin{abstract}

Dark matter simulations require statistical techniques to properly identify and classify their halos and structures. Nonparametric solutions provide catalogs of these structures but lack the additional learning of a model-based algorithm and might misclassify particles in merging situations. With mixture models, we can simultaneously fit multiple density profiles to the halos that are found in a dark matter simulation. In this work, we use the Einasto profile \citep{1965TrAlm...5...87E, 1968PTarO..36..414E, 1969Afz.....5..137E} to model the halos found in a sample of the Bolshoi simulation \citep{2011ApJ...740..102K}, and we obtain their location, size, shape and mass. Our code is implemented in the R statistical software environment and can be accessed on \texttt{https://github.com/LluisHGil/darkmix}.

\end{abstract}


\vfill 

\section{Introduction}
\label{sec:intro} 

\subsection{Dark Matter Halos}

In 1952, Neyman \& Scott proposed the first statistical model of large-scale galaxy distribution \citep{1952ApJ...116..144N, 1953PNAS...39..737N, 1954PNAS...40..873N}. This model interpreted luminous matter to be distributed in the universe according to a stochastic process, where a discrete distribution of galaxies is aggregated into overdense clusters that themselves are distributed in space as a Poisson process. This model needs three main components to be built: the distribution of galaxies within each cluster, the size distribution of the clusters, and a description of the clustering of clusters. While their original formulation was too simple, this framework is still a valid approach for the baryonic distribution in the universe and it can be extended to dark matter distribution \citep{2002PhR...372....1C}. Here, dark matter is a continuous field of particles that collapses from overdensities into strongly clustered structures, which are called halos. The identification and morphology of these halos is the main topic of the present work. 

Analytic models and numeric simulations show how the initial dark matter field, which is made of particles, evolves from an initially smooth state to a highly clustered final condition, which results in a complex cosmic web of knots (the halos), filaments, sheets and voids \citep{1985ApJS...58....1B}.  Simulations show that the halo mass \citep{1999MNRAS.310.1147M, 1996ApJ...462..563N}, abundance and spatial distribution \citep{1997ApJ...484..523C, 2001MNRAS.321..372J}  are highly dependent on the initial conditions.  The final structure within a halo can be reasonably assumed to be in virial equilibrium.  The nature and evolution of the galaxies that have formed inside halos is strongly dependent on the parent halo's properties \citep{1999MNRAS.303..188K, 1999MNRAS.310.1087S, 2001MNRAS.327.1041B, 2000MNRAS.319..209C}, with more galaxies formed in more massive and clustered halos.  However, the galaxy distribution is biased toward stronger clustering conditions \citep{2003MNRAS.344..847M, 2011ApJ...736...59Z, 2016ApJ...818..174H}.

The spherical collapse model is a classic approximation for initial conditions leading to dark matter halos \citep{1972ApJ...176....1G, 1984ApJ...281....9F, 1985ApJS...58....1B}. Here, dark matter overdensity collapses from a tophat density perturbation into a halo that, depending on the overdensity mass and density,  virializes when a certain size is reached. The final density of the halo is much higher that the prediction of a linear model because the evolution of the halo clustering is governed by nonlinear processes, where the densest regions are populated by the most massive halos \citep{1980PhRvD..22.1882B, 1984ApJ...284L...9K} and the dark matter follows a log-normal distribution \citep{1991MNRAS.248....1C, 2017A&A...601A..40H, 2017MNRAS.466.1444C}.

The shape of the collapsed halos, \cite{1985ApJS...58....1B} and \cite{1984ApJ...281....9F} suggests that their density profile around the center depends on the initial density distribution of the parent overdense region. Although more massive halos arise from denser peaks in the initial fluctuation field \citep{1984ApJ...284L...9K, 1985ApJ...297...16H}, these dense peaks are also less centrally concentrated \citep{1980PhRvD..22.1882B}.  Massive virialized halos are thereby less centrally concentrated than less massive halos \citep{1996ApJ...462..563N}. 

Several halo density profiles have been proposed, including the \citet{1990ApJ...356..359H}, Navarro-Frenk-White \citet{1996ApJ...462..563N, 1997ApJ...490..493N}, and \citet{1968PTarO..36..414E} profile. In sections~\ref{sec:NFW} and~\ref{sec:einasto}, we introduce the two later profiles and we justify our final selection of the Einasto profile.

\subsection{Finding Structure with Mixture Models and Other Methods}

The growth in volume and detail of astronomical observations and simulations requires automated tools to characterize the properties and evolution of halo structures. These tools should be robust and objective, and should not depend on heuristic choices or subjective judgment.

Most of the widely-used clustering algorithms for galaxy clustering analysis are nonparametric and are based on dissimilarity distances, which is a metric that is used to decide if two particles are sufficiently close to belong to the same cluster.  Astronomers commonly use the Friends-of-Friends algorithm, which is known in statistics as single-linkage hierarchical agglomerative clustering \citep{gower1969minimum}. This method is highly performant and is scalable, which is crucial for large volume data sets, such as dark matter numerical simulations \citep{2020arXiv200311468W}. However, this method is prone to `chaining' of unrelated groupings \citep{everitt2011cluster}. The resulting clusters depend strongly on the choice of density or size threshold value. Other nonparametric clustering procedures---such as Ward's hierarchical clustering, $k$-means algorithm, DBSCAN, kernel density estimation bump hunting, and their many extensions---also have heuristic thresholds or stopping rules with a corresponding loss of objectivity and robustness.  These procedures also have the limitation of giving `hard' classifications where data points are strictly classified on one cluster or another without regard to the reliability of this decision. Therefore, the characterization of merging or blended clusters is limited. 
In contrast, parametric methods assume a particular shape to the structures in the population.  They also have the advantage that modeling can be based on maximum likelihood estimation or (if prior information is available) Bayesian inference, without requiring the addition of heuristic thresholds or stopping rules. These methods are typically based on probability density functions. This allows us to perform statistical tests (e.g., significance tests on the parameters) and goodness of fit validations (e.g., the coefficient of determination) \citep{rao1973linear}. Furthermore, they give `soft' probabilities for each point belonging to each cluster. Hard membership classifications can be decided afterwards using heuristic decision rules. 

These benefits motive us to use finite mixture models to detect and characterize dark matter structures \citep{peel2000finite, mclachlan2000mixtures, everitt2005finite, fruhwirth2006finite, mclachlan2007algorithm, everitt2011cluster}. This technique has been widely used in astronomy and astrophysics with considerable success \citep{2014ApJ...787..107K,  fruhwirth2019handbook, KuhnFeigelson19}.

We understand the dark matter distribution to be structured in halos, which can be described by a parametric surface-density distribution. The halos and other structures are described by the mixture model components, which are summed to create the surface density function. Mixture models are widely used in parametric modeling of point processes, usually with Gaussian function components \citep{fraley2002model}. As previously explained, we will instead use an astrophysically motivated function, such as the Einasto profile.

The mixture model is then estimated following a three step method: first, the number of components is chosen by the user; second, the properties of these components are obtained through maximum likelihood parameter estimation (MLE); and third, the component assignments for dark matter particles are determined using posterior probabilities from the fitted models \citep{everitt2011cluster}.

Mixture models are dependent on three main choices: the chosen number of components, the convergence criteria towards the surface density distribution, and the selected profile. The first is assessed using model selection, such as Bayesian and Akaike Information criteria \citep{schwarz1978estimating, akaike1998information}. The model goodness-of-fit will be measured using well-known statistics, such as the coefficient of determination, or residual maps between the input data and model predictions \citep{2014ApJ...787..107K, 2017MNRAS.472.2808D}. We will justify the election of the Einasto profile in section~\ref{sec:einasto}. Once the model is estimated, we may want to give a final particle classification in the  clusters. Mixture models allow for a probability based classification, but as we explain in section~\ref{sec:mem} we recommend the application of a heuristic threshold for background particles. This threshold should be based on the merging conditions of our data.

Aside from the mixture model alternatives, several algorithms are widely used by the community to detect structures in dark matter N-body simulations and classify their particles. We mention the Bound Density Maximum algorithm \citep{1997astro.ph.12217K, 2013AN....334..691R} in section~\ref{bdm}, which is based on density maximums and spherical halos. Another alternative is the ROCKSTAR algorithm \citep{2013ApJ...762..109B}, `based on adaptive hierarchical refinement of friends-of-friends groups in six phase-space dimensions and one time dimension'. With this method, the algorithm can provide an effective particle classification and can even detect small subhalos in merging conditions. The halo finder VELOCIraptor \citep{2019PASA...36...21E} has also been proven to be able to `identify sub-halos deep within the host that have negligible density contrasts to their parent halo'. Both methods are meant to be used on large N-body simulations, providing a full catalog of halos, sub-halos and even tidal features.

In contrast, our method is a parametric based approach. Although this limits its use to small-sized samples of particles, it provides a parametric fitting of the halos density profile. In this work, we will focus on the advantages of obtaining such a profile-based description.

This paper is organized as follows.  Section~\ref{sec:mm} presents the finite mixture modelling technique, the Einasto profile and different applications for the estimated results. Section~\ref{sec:mle} illustrates the MLE calculation and section~\ref{sec:ra} describes the tools that we used to validate the best-fit model. In section~\ref{sec:over} we summarize the steps of our code \texttt{darkmix} and we apply it in section~\ref{sec:val} to a set of generated realizations of a simulated dark matter distribution with Einasto profile and fit them with our software to validate it. Section~\ref{sec:data} presents a data set from the Bolshoi simulation, and section~\ref{sec:res} shows the analysis and results of our mixture modeling and validation. In section~\ref{sec:con}, we summarize our main conclusions and outline future work.   

\section{Finite Mixture Models for Dark Matter Halos}
\label{sec:mm}

Finite mixture densities are a family of probability density distributions that combine multiple components into a single probability function. Each component has a probability density function (e.g., the Einasto profile) and the final mixture model is the weighted sum of $c$ components. All components are continuous, and can be evaluated at any location of the data space, occupied or not by a particle. Mixture models admit as many different components as desired, as long as they can be defined as probability distributions. In this work we will consider two kinds of components: $k$ halos, and a single background component containing all of the particles that are not associated with a halo. The total number of the mixture model is $c = k + 1$.

Given a data sample of $N$ dark matter particles from a simulation, we define $\mathbf{X}$ as a three column matrix containing the coordinates of our particles. Each component will be modeled by a profile function $\rho_j(\mathbf{r}_i, \vec{\theta}_j)$, $j=1,\dotsc,c$, where $\vec{\theta}_j$ is the parameters vector for component $j$ and $\Theta$ is the matrix of the $c$ vectors $\vec{\theta}_j$. The sum of these components is weighted by the mixing proportions $\vec{w} = \{w_j\}$, $j=1,\dotsc,c$, which are non-negative. In point process statistics, the model is defined over a window $W$ containing the sample of $N$ points $\mathbf{X} = \{\mathbf{r}_i\}$, where $\mathbf{r}_i = (x_i, y_i, z_i)$. 
Together, the finite mixture model $\Sigma$ can be written as

\begin{equation}\label{fun:mm}
\Sigma(\mathbf{X} | \vec{w}, \Theta) = \sum_{j=1}^{c} w_j \cdot \rho_j(\mathbf{X} | \vec{\theta}_j) = \sum_{j=1}^{c} \sum_{i=1}^N w_j \cdot \rho_j(\mathbf{r}_i | \vec{\theta}_j)
\end{equation}

Our implementation of the mixture model for the dark matter distribution will follow the strategy of \citep{2014ApJ...787..107K}, and therefore we chose our notation after this paper.

Functions $\rho_j$ represent the profile functions of our dark matter structures, which are summed into the surface density function $\Sigma$ and the probability density function that we use as a model and will fit to our data. These functions typically create a multimodal distribution that matches the clusters that are present in our data. Therefore, individual dark matter particles are used to obtain the overall density distribution of the structure that they belong to. No learning can be obtained from the internal distribution (i.e., the relative position of the particles inside their structure). The total finite mixture model $\Sigma$ is the weighted sum of these components and models the dark matter density distribution as generated by the N-body simulation.

The profile of a mixture model component can be as irregular as we are able to model it. However, since this method is generally used as a cluster classification method, we will model two different components---the halos and the background.

We will introduce the NFW and the Einasto profiles in the next section. However, as we justify, we will only make use of the latter. We will also introduce the final version of our mixture model before deducing several quantities and functions of interest.

\subsection{The NFW Profile}\label{sec:NFW}

The NFW profile \citep{1996ApJ...462..563N, 1997ApJ...490..493N} has the following shape

\begin{equation}
  \rho(r|M) = \frac{\rho_s(M)}{\left[ c(M) \frac{r}{r_{\rm vir}(M)} \right] \left[ 1 +  c(M) \frac{r}{r_{\rm vir}(M)} \right]^2} \; ,
\end{equation}

truncated at the virial radius $r = r_{\rm vir}(M)$.
This profile has a logarithmic slope of $-1$ at small scales ($r \ll r_{\rm vir}/c$) and of $-3$ at large scales ($r \gg r_{\rm vir}/c$).
Here, $\rho_s(M)$ is a normalization factor that is fixed from the condition that the total mass integrated to $r_{\rm vir}(M)$ must be equal to $M$ and $c(M)$ is the concentration parameter. We note that the NFW profile or power-law profiles cannot be used in this context because they do not have a finite integral. Instead, we will use the Einasto profile for our study.

\subsection{The Einasto Profile}\label{sec:einasto}

The Einasto profile \citep{1965TrAlm...5...87E, 1968PTarO..36..414E, 1969Afz.....5..137E} for three-dimensional density profiles is similar to the S\'ersic profile used for two-dimensional galaxy brightness profiles \citep{1963BAAA....6...41S,1968adga.book.....S}. Assuming spherical symmetry, the Einasto profile describes the density $\rho$ of halo $j$ as a function of distance from the halo center $\mathbf{r_{0,j}}$. The S\'ersic index $n$ is a free parameter that is used to measure the shape of the halo density profile: larger values of $n$ create centrally concentrated profiles. As we will see in section~\ref{sec:res}, large values of this parameter (above $n \sim 8$) degenerate the profile into a power law-like function. The other free parameter of the profile is Einasto's radius $r_e$, which defines a volume containing half of the total mass that can be used to understand the size of the halo.

Given these parameters, we can define $d_n$, which is a function of $n$ such that $\rho_e$ is the density at the radius $r_e$, and we have defined $\rho_e$. The factor $d_n$ can be obtained by solving

\begin{equation}
\Gamma(3n) = 2\gamma(3n,d_n)
\end{equation}

\noindent where $\Gamma$ is the complete gamma function and $\gamma$ is the lower incomplete gamma function\footnote{The gamma functions are extensions of the factorial function to complex numbers.}.

We are now ready to define the Einasto profile as

\begin{equation}
\rho(r) = \rho_e \exp{\Big(-d_n \Big[ (r/r_e)^{1/n} - 1\Big]\Big)}
\end{equation}

\noindent where $r = ||\mathbf{r-r_{0}}||$ is the distance between a chosen location $\mathbf{r}$ and the center of the halo $\mathbf{r_0}$. As we see in eq.~\ref{fun:mm}, this profile will be multiplied by parameter $p_j$, which is the mixture coefficient. This makes it impossible for us to estimate $p_j$ and $p_e$ separately. The results that we provide for $p_j$ in Tables~\ref{dens} and~\ref{tab:6k} should be read as $p_j \cdot p_e$.

Before moving on to the next section, we define the profile of the background component as a constant density, having value 1 at all locations of the window $W$. Since the background profile is multiplied by its our mixture coefficient $p_b$, we expect it to be close to the mean density of the data set. In a large data set, where conditions are comparable to the universe, we would expect it to be around the mean density of the universe. This is equivalent to defining the background as a homogeneous Poisson distributed point process, with profile function

\begin{equation}
\rho_b(\mathbf{r}) = 1
\end{equation}

\subsection{Dark Matter Halo Mixture Model}

The mixture model for our problem is a weighted sum of the background component ($\rho_b$) plus the $k$ halo components ($\rho$), hence $c = k + 1$ components (see eq.~\ref{mmsigma}). The parameters for each halo profile are the three vectors representing the halo center $\mathbf{r_0}$, the size parameter $r_e$, and the shape parameter $n$. These parameters  are collected into a five-dimensional parameter vector $\vec{\theta} = (\mathbf{r_0},r_e,n)$. 
 
The contribution of each component to the final density distribution is uneven. The mixture proportions $\vec{w}=\{w_j\}$, $j=1,\dotsc,c$ are used as weights to normalize each halo's contribution to the mixture model. Based on equation (\ref{fun:mm}), the resulting model is 

\begin{equation}\label{mmsigma}
\Sigma(\mathbf{r} | \vec{w},\Theta) = \frac{N}{M} \Big(w_{b} \cdot \rho_b(\mathbf{r})  + \sum_{j=1}^{k} w_j \cdot \rho(\mathbf{r}-\mathbf{r_{0,j}} | r_{e,j},n_j) \Big)
\end{equation}

\noindent  where the weights $w_j$ give the mixture proportions $p_j = N\cdot w_j/M$ in equation~(\ref{fun:mm}) and $M$ is the total mass given by

\begin{equation} \label{mmmass}
M = \int_W \Big(w_{b} + \sum_{j=1}^{k} w_j \cdot \rho(\mathbf{r}-\mathbf{r_{0,j}} | r_{e,j},n_j) \Big) d\mathbf{r}
\end{equation}

The term $N/M$ works as a normalizing constant so that the integral of the model $\Sigma$ is always the number of particles $N$. Note that we ignore function $\rho_b(\mathbf{r})$ because it is constant 1. Function $\Sigma$ is our probability density function for a mixture model problem as in equation~\ref{fun:mm}. Since the density of the universe is constrained to not be infinite, we can model the density of the components relative to each other. In equation \ref{mmsigma}, the mixture proportions are defined so that $\sum_{i=1}^c p_i = N$, and by definition we can set $w_1 = 1$.

This normalization of the model by the total number of objects $N$ has other advantages. The integral of $\Sigma(\mathbf{r})$ in a region $A \subset W$ gives the number of model objects in $A$. The integration of a chosen model component over the entire volume $W$ gives the number of model objects belonging to this component. The statistical model can thus be understood as an inhomogeneous Poisson distribution with intensity $\Sigma(\mathbf{r})$. 

Notice how the quantity $\rho_e$ in the Einasto profile will be masked into the mixture proportion  as a consequence of the mixture model because both are factors to the profile.

\subsection{Halo Identification and Characterization}
\label{sec:su}

Once the mixture model is fitted (section~\ref{sec:mle}), analysis of each Einasto-shaped component representing a halo can follow. Due to the additive nature of a mixture model, it is easy to determine the dominant component at every location. Here, we apply a simple decision rule to assign individual data points to a halo: given a particle, we calculate its probability to belong to each component. We then randomise the membership of the particle using the probabilities.

Each model halo component can be compared to the empirical halo profile in the context of the total mixture model of overlapping halos.  Component $j$ has a three-vector $\mathbf{r_{0,j}}$, giving its location, and three Einasto parameters $w_j$,  $r_{e,j}$ and $n_j$.  We define the empirical halo profile as the number of real particles per volume, the particle density, in  concentric shells centred at $\mathbf{r_{0,j}}$ with volumes $S_j(r) = V_j(r + \text{d}r) - V_j(r)$.  The empirical halo profile is

\begin{equation}\label{pro:emp}
    \hat{\delta}_{j}(r) = \frac{1}{S_j(r)}(n_j(r+\text{d}r)-n_j(r))
\end{equation}

\noindent where $n_j(r)$ is the number of particles in a sphere of radius $r$ and center $\mathbf{r_{0,j}}$.  Note this has the same units, number of particles per unit volume, as the mixture model in equation~\ref{mmsigma} by $N/M$. The estimated density profile of the mixture model centered at $\mathbf{r_{0,j}}$ is 
\begin{equation}\label{pro:fit}
    \hat{P}(r | \mathbf{r_{0,j}}) = \int_{S_j(r)} \Sigma(\mathbf{r} | \vec{p}, \Theta) d\mathbf{r}.
\end{equation}

The integral of $\Sigma$ at any volume $V$ gives us the estimated number of particles. Since profile $\hat{P}(r)$ is evaluated at the concentric shells $V_j(r)$ with thickness $\text{d}r$, it should be understood as a density profile over radius $r$. Component $j$ in the previous integral can be isolated from the total model profile as  

\begin{equation}\label{pro:com}
    \hat{\rho}(r | \mathbf{r_{0,j}}) = \int_{S_j(r)} p_j\rho(\mathbf{r - r_{0,j}} | r_{e,j}, n_j) d\mathbf{r}.
\end{equation}

We can now compare the observed profile of $\hat{\delta}_{j}$ with the full model estimator $\hat{P}(r | \mathbf{r_{0,j}})$ and  with the isolated estimated profile $\hat{\rho}(r | \mathbf{r_{0,j}})$, which may show important departures from the total profile.  For short distances, $\hat{\rho}$ and $\hat{P}$ should have similar values. However, for distances at which our component $j$ starts to increasingly overlap with other components, these functions will start to diverge as other structures contribute more to $\hat{P}$. 

Parametric models offer the possibility of easily generating new samples following the model distribution. For a mixture model, this is done with an inverse transform sampling for the estimated number of objects per component $N_j$.

\begin{equation}\label{npart}
    N_j = \int_W p_j\rho(\mathbf{r - r_{0,j}} | r_{e,j}, n_j) d\mathbf{r}.
\end{equation}

For the possible interest of the user, software implementations of these functions are included in the $R$ language and are included in our repository.

\section{Model Fitting} \label{sec:fitting}

\subsection{Estimation of Model Parameters and Model Selection}\label{sec:mle}

The optimal mixture model for a data set is calculated by maximum likelihood estimation (MLE) for the log-likelihood

\begin{equation}\label{loglikn}
\log {L(\vec{p},\Theta | \mathbf{X})} = \sum_{i=1}^N \log {\Sigma(\mathbf{r}_i | \vec{p},\Theta)} - \int_W \Sigma(\mathbf{r'} | \vec{p}, \Theta) d\mathbf{r}'
\end{equation}

\noindent where $\mathbf{X} = \{\mathbf{r}_i\}$ contains the point process distributed in the window $W$ with surface density distribution $\Sigma$. Note that the right-hand side term is the mass $M$ for the parameters $\Theta$. The MLE and Bayesian best-fit model parameters $\Theta$ are calculated for a chosen number of $c$ components. Model selection among models of different complexities is based on minimising two commonly-used penalized likelihood measures,  the Bayesian Information Criterion (BIC) and the Akaike Information Criterion (AIC) \citep{schwarz1978estimating, akaike1998information}, 
\begin{align}
\text{BIC}(k) & = -2 \log L + 6(c-1) \log N \label{eq:bic}\\
\text{AIC}(k) & = -2 \log L + 12(c-1) \label{eq:aic}
\end{align}
\noindent where $6(c-1)$ is the number of parameters in $c-1$ halo components plus the background component and $\text{log} L$ is the log-likelihood for the best fit parameters.  

The relative strengths of AIC and BIC is widely debated, although both are founded on powerful theorems \citep{lahiri2001model, konishi2008information, burnham2002model, kass1995reference, everitt2011cluster}.  The BIC has a well-accepted valuation for relative model merit: one model is strongly (very strongly) favored over another when $\Delta(BIC) > 6$ ($>10$) \citep{kass1995bayes}.  For the mixture model problem, \citet{2014ApJ...787..107K} found that the AIC was more sensitive to the presence of sparse clusters in the presence of rich clusters because its penalty for complexity is weaker when $N$ is large.

\subsection{Goodness-of-Fit} \label{sec:ra}

A best-fit model with optimum complexity selected with a likelihood-based criterion is not guaranteed to be a good fit to the data. A complex clustered spatial distributions cannot be effectively fitted with a mixture model of a few halos. It is therefore necessary to assess the overall quality of the fit for the entire pattern by a study of the residuals to identify departures of the model from the real data density distribution. Several such tests are outlined here. Studies involving residual analysis of astronomical spatial point processes for goodness-of-fit evaluation of maximum likelihood mixture models include \citet{2014ApJ...787..107K} and \citet{2017MNRAS.472.2808D}.

The residual analysis that is used here is described in \citet{RSSB:RSSB519, baddeley2015spatial} as a `raw residuals'. Raw residuals are defined as the absolute difference between the real number of points in a region $A$ and our estimation for the same region. For our mixture model,

\begin{equation}\label{res}
R(A) = n(\mathbf{X} \cap A) - \int_A \Sigma(u | \vec{p}, \Theta) \text{d}u
\end{equation}

\noindent where $n(\mathbf{X} \cap A)$ is the number of data points in the region $A$. For well-fitted models, the sum of the raw residuals should approach zero when integrated over $W$, and the residual map should approach a random spatial distribution with no correlations between the values of the residuals and the locations of the data points (spatial white noise). Residuals should have low amplitude when compared with the surface density function based on the data. 

The calculation of the residuals is made using a quadrature or grid of dummy points $\mathbf{Q} = \{u_i\} \subset W$, $i=1,\dotsc,T$ \citep{RSSB:RSSB519}. Each point defines a small cell where the residuals will be calculated. These residuals create a sparse distribution that is 1 when the cell contains a data point $\mathbf{r}$ and $-\Sigma(u | \mathbf{X}, \vec{p}, \Theta)$ at empty locations in $W$, which is negative and typically close to zero.

To effectively visualise the spatial distribution of the residuals, they have to be smoothed. With grid $\mathbf{Q}$ dense enough to approximate an integral, we can obtain the smoothed residual map $s(u)$ with

\begin{equation}\label{eq:su}
s(u) = \int_W \kappa_{\omega}(u-v) \text{d}R(v) = \sum_{i=1}^N \kappa_{\omega} (u - \mathbf{r}_i) - \int_W \kappa_{\omega}(u-v) \cdot \sum (u | \mathbf{X}, \vec{p}, \Theta) \text{d} v  
\end{equation}

\noindent where $\kappa_{\omega}$ is a kernel function and $R_{\Theta}(v)$ is the raw residual in a cell $v$. 

It is also useful to define the relative residuals $e(u)$, which are defined as the residual map normalized by the model intensity. The model intensity $\Sigma^{\dag}_{\omega}$ is the right-hand term of eq.~\ref{eq:su}. Hence the definition of these two functions is

\begin{equation}\label{model}
\Sigma^{\dag}_{\omega}(u | \vec{p}, \Theta) =  \int_{W} \kappa_{\omega}(u-v) \Sigma(u | \vec{p}, \Theta) \text{d}v
\end{equation}

\begin{equation}\label{relerr}
e(u) = s(u)/\Sigma^{\dag}_{\omega}(u | \vec{p}, \Theta)
\end{equation}

As we will see in the following sections, this function can be used to detect data structures that have not been modeled by any model component. For any structure that is properly mapped by the model, $s(u)$ will be a small quantity and the values of $e(u)$ in its region will also be small. However, if a structure in the data is not included in the model, then the value of function $\Sigma^{\dag}_{\omega}(u | \vec{p}, \Theta)$ will be close to zero for any location close to that structure and the error value in $s(u)$ will be amplified in $e(u)$ and unfitted structures can be easily detected.

The kernel function that is used in this work is the Gaussian filter, which is commonly used in cosmology \citep{2002sgd..book.....M}, where $\omega$ is the smoothing radius or bandwidth. The appearance of the smoothing strongly depends on the choice of this quantity. The bandwidth can be selected using cross-validation or other techniques and we select it heuristically to give informative residual maps.   

Finally, we use the coefficient of determination \citep{rao1973linear} to assess the global goodness of fit. Using the expected proportionality between the model density ($\Sigma_{\omega}^{\dag}$) and the data density ($\Sigma_{\omega}^{*}$) functions, we estimate a simple linear regression between them and use the resulting $R^2$ coefficient as a measure of goodness of fit \citep{2017MNRAS.472.2808D}.

The data $X$ can be similarly smoothed with the same kernel. This function is not used in the fitting or model validation process, which would be incorrect given the loss of data structure detail. However, we include it in our code repository for completeness and visualization purposes (as in Fig~\ref{data_model_dens} top left-hand panel):

\begin{equation}\label{ker}
\Sigma^*_{\omega}(u) = \sum_{i=1}^N \kappa_{\omega}(u-\mathbf{r}_i) \qquad \mathbf{r}_i \in W \\
\end{equation}

\subsection{Available Software Packages}

It is not our aim to compare the performance of our code with other packages available for users. In contrast, our solution is adapted to the particular scenario of dark matter halos with Einasto profile, while most of the public solutions use GMM or only accept two-dimensional data. However, we find it interesting to provide a small summary of the available solutions.

Substantial code devoted for mixture model analysis are available in the R public domain statistical software environment such as CRAN packages {\it mixtools}, {\it mclust}, and {\it EMCluster} \citep{mixtools2009, mclust16, Chen2015EMClusterpackage}. The Python Machine Learning library \textit{scikit-learn} \citep{scikit-learn} includes Gaussian mixture models (GMM) algorithms. Another option is $EMMIX$, which is written in Fortran \citep{RePEc:jss:jstsof:v:004:i02}, where the Expectation-Maximization (EM) algorithm is used to find the MLE \citep{krishnan1997algorithm,mclachlan2008algorithm}. However, estimation of best fits in complex data sets can be difficult to achieve thanks to the multiple peaks. A more robust procedure makes use of the stochastic EM algorithm \citep{celeux1992classification}, where randomization of the steps seeks to avoid trapping in the first found local maximum. Even with this technique, it is advisable to repeatedly run estimation algorithms using different starting values to avoid convergence to a non-optimal local maximum. If the same best fit solution is achieved every time, then we can be more confident of reaching an absolute maximum.

\subsection{Software Repository}

Our solution necessarily departs from the previous packages. As explained, while most mixture model software assumes Gaussian shapes and use the EM algorithm for parameter estimation, we will adopt different algorithms. Mixture model solutions with non-Gaussian functions such as the Einasto profile are not so easily achieved with this method, due to the reduced curvature in the derivative. 

Following \citet{2014ApJ...787..107K}, who calculate two-dimensional mixture models with the isothermal ellipsoid shapes, we will maximize the log-likelihood function using the Nelder-Mead simplex algorithm \citep{nelder1965simplex} as implemented in function \textit{optim} within CRAN package \textit{stats}  \citep{team2015r}. Testing with different initial values is recommended to avoid trapping in local maxima.  Since the models have high dimensionality, a strategy of systematic freezing and thawing parameters during estimation can be useful. Once the Nelder-Mead algorithm has finished a first attempt at estimation, the results can often be improved by freezing some of the parameters and repeating the calculation for the remaining free parameters. This technique is valuable in mixture models because the parameters of distant halos tend to be uncorrelated. It can also be helpful to obtain a good estimates of halo centers $\mathbf{r_0}$ before attempting the fitting of the rest of the parameters.  Confidence intervals on MLE parameters can be estimated from the Fisher Information Matrix.  

MLE can present additional problems that depend on the profile functions $\rho_i(\mathbf{X},\Theta)$. It is possible for singularities to exist for certain values of $\mathbf{X}$ or $\Theta$ where the likelihood becomes infinite. This might happen when the number of parameters to be estimated is high when compared with the sample size. This problem can be addressed with a Bayesian approach where the posterior distribution of the parameters is mapped instead of maximizing the log-likelihood function. As priors, we use Gaussian functions with large variances for the centers of the halos, and log-normal distributions for the size and shape parameters $r_e$ and $n$. These calculations are performed using Markov chain Monte Carlo (MCMC). However, even if aided by MLE results that were previously obtained by the Nelder-Mead algorithm, sufficient  MCMC evaluations of high-dimensional parametric models can be computationally expensive. 

The software that we have used for our Bayesian calculations is the CRAN package \textit{LaplacesDemon} \citep{LAP1,LAP2,LAP3,LAP4}, which provides more than 40 different MCMC algorithms. These algorithms make different decisions regarding the next combination of parameters $\Theta$ to be tested to efficiently map the normalized log-likelihood function~\ref{loglikn}. Adaptive MCMC algorithms, which use the previous evaluations to choose the next $\Theta$, are often more efficient in finding the overall distribution of the function but we must always finish with a long run of a non adaptive algorithm to ensure convergence. In this work, we make use of the \textit{Adaptive Metropolis-Within-Gibbs} (AMWG, \cite{doi:10.1198/jcgs.2009.06134}) and the \textit{twalk} \citep{christen2010general} algorithms.  These calculations start with a preliminary MLE solution that is obtained with the Nelder-Mead algorithm.

Our software implementation of this mixture model decomposition of the dark matter distribution is named {\it DarkMix} and is written in the R statistical software environment, which has many tools relating to spatial point processes, mixture models, likelihood calculations, model assessment, and graphics. Our work extends that developed by \citet{2014ApJ...787..107K} to identify star clusters in two dimensions. We present a new library to model three-dimensional data sets \citep{darkmix_zenodo}, which is publicly available on Github\footnote{https://github.com/LluisHGil/darkmix \label{github.url}}.  
Its use is explained in sections~\ref{sec:over} and~\ref{sec:res}.
Additional documentation for the code can be found online\footnote{https://darkmix.readthedocs.io/}.

\section{\texttt{Darkmix} Overview}\label{sec:over}

With the equations from section~\ref{sec:fitting} the \texttt{darkmix} code has the following capabilities:

\begin{enumerate}
    \item Create data and model objects that are compatible with the R library \texttt{spatstat}. This is a powerful library to work with point processes and its capabilities can be combined with our code.
    \item Parametric estimation with two available methods: the Nelder-Mead simplex algorithm \citep{nelder1965simplex} and MCMC.
    \item Functions are incorporated for model selection (through AIC and BIC calculations) and goodness-of-fit (through $R^2$ and residuals plots).
    \item Generation of model outputs, which comprise the soft classification of the particles, the extraction of individual profiles for components, the generation of model realizations and the visualization of the components' sizes.
\end{enumerate}

\section{Model Validation}\label{sec:val}

Before we introduce our real case data set in section~\ref{sec:data} and estimate a mixture model, we must first validate the performance of our algorithm. 
To do so, we will estimate the mixture model in three different configurations: 3, 9 and 16 dark matter halos with background particles in a cube of 25 length units side. For each configuration, we generate 20 realizations with three different particle densities: $0.25$, $0.5$ and $0.75$ particles per volume unit. 

The halos are modeled using the Einasto profile and the background components is uniformly distributed, as assumed by our model. The true parameters defining the location, size and shape of the halos, together with the number of particles, are generated randomly (see Tables~\ref{tab:val9} and~\ref{tab:val16}). We then obtained the mixture coefficients using eq.~\ref{mmmass} knowing that $w_0 = 1$. When using three halos, the configuration is a simple set of components around the center of the volume, and its fitting presented no problem. We will use the more complex cases of 9 and 16 halos to study merging halos or locate halos near the volume window boundary.  

\begin{deluxetable}{lrrrrrrr}
\caption{True parameters for model validation: 9 halos} \label{tab:val9}
\tabletypesize{\small}
\tablehead{
\colhead{k} & \colhead{$x_0$} &\colhead{$y_0$} &\colhead{$z_0$} &\colhead{$r_e$} &\colhead{$n$} &\colhead{$\log w$} &\colhead{$N$}  }
\startdata
$1$ & $2.9$ & $21.0$ & $21.7$ & $1.1$ & $2.4$ & $0.0$ & $256$ \\
$2$ & $8.2$ & $6.5$ & $18.6$ & $0.9$ & $1.4$ & $0.68$ & $544$ \\
$3$ & $8.7$ & $14.9$ & $16.0$ & $1.4$ & $1.7$ & $-0.86$ & $66$ \\
$4$ & $10.1$ & $16.2$ & $4.4$ & $1.3$ & $1.9$ & $-0.64$ & $92$ \\
$5$ & $16.1$ & $7.7$ & $5.9$ & $2.0$ & $1.7$ & $-0.42$ & $518$ \\
$6$ & $16.4$ & $22.8$ & $19.5$ & $2.2$ & $2.9$ & $-0.65$ & $454$ \\
$7$ & $18.4$ & $16.3$ & $22.3$ & $1.4$ & $1.9$ & $-0.08$ & $403$ \\
$8$ & $20.3$ & $6.1$ & $13.9$ & $0.7$ & $1.7$ & $1.09$ & $717$ \\
$9$ & $21.7$ & $14.9$ & $7.9$ & $1.1$ & $2.3$ & $0.20$ & $415$ \\
\enddata
\tablecomments{These parameters have been randomly generated in a cube of side 25 to describe a configuration of 9 Einasto halos (here they are sorted by $x_0$) with density 0.25 particles per volume unit. The background contains 442 particles, which gives $\log w_b = -2.4$. The populations $N$ for densities 0.5 and 0.75 can obtained multiplying the values in the table by 2 and 3.}
\end{deluxetable}

\begin{deluxetable}{lrrrrrrr}
\caption{True parameters for model validation: 16 halos} \label{tab:val16}
\tabletypesize{\small}
\tablehead{
\colhead{k} & \colhead{$x_0$} &\colhead{$y_0$} &\colhead{$z_0$} &\colhead{$r_e$} &\colhead{$n$} &\colhead{$\log w$} &\colhead{$N$}  }
\startdata
$1$ & $1.3$ & $6.2$ & $5.2$ & $1.3$ & $1.9$ & $0.0$ & $204$ \\
$2$ & $2.9$ & $21.0$ & $21.7$ & $1.1$ & $2.4$ & $-0.21$ & $83$ \\
$3$ & $5.4$ & $15.7$ & $5.2$ & $1.1$ & $1.8$ & $0.76$ & $729$ \\
$4$ & $5.7$ & $20.0$ & $6.9$ & $2.0$ & $2.1$ & $-0.57$ & $212$ \\
$5$ & $7.9$ & $15.4$ & $12.8$ & $1.0$ & $1.4$ & $0.33$ & $179$ \\
$6$ & $8.2$ & $6.5$ & $18.6$ & $0.9$ & $1.4$ & $0.25$ & $108$ \\
$7$ & $8.7$ & $14.9$ & $16.0$ & $1.4$ & $1.7$ & $-0.56$ & $69$ \\
$8$ & $9.2$ & $21.3$ & $19.4$ & $0.9$ & $1.6$ & $0.29$ & $130$ \\
$9$ & $10.1$ & $16.2$ & $4.4$ & $1.3$ & $1.9$ & $-0.58$ & $55$ \\
$10$ & $15.5$ & $10.6$ & $5.7$ & $1.0$ & $1.8$ & $0.71$ & $481$ \\
$11$ & $16.1$ & $7.7$ & $5.9$ & $2.0$ & $1.7$ & $-0.4$ & $293$ \\
$12$ & $16.4$ & $22.8$ & $19.5$ & $2.2$ & $2.9$ & $-0.84$ & $154$ \\
$13$ & $18.4$ & $16.3$ & $22.3$ & $1.4$ & $1.9$ & $-0.5$ & $83$ \\
$14$ & $20.3$ & $6.1$ & $13.9$ & $0.7$ & $1.7$ & $0.91$ & $255$ \\
$15$ & $21.7$ & $14.9$ & $7.9$ & $1.1$ & $2.3$ & $0.22$ & $230$ \\
$16$ & $23.8$ & $15.5$ & $16.8$ & $1.2$ & $2.3$ & $0.55$ & $571$ \\
\enddata
\tablecomments{These parameters have been randomly generated in a cube of side 25 to describe a configuration of 16 Einasto halos (here they are sorted by $x_0$). Components 2, 6, 7, 9, 11, 12, 13, 15, and 16 are copied from the 9 halos configuration (see Table~\ref{tab:val9}). The background contains 68 particles, which gives $\log w_b = -2.85$. The populations $N$ for densities 0.5 and 0.75 can obtained multiplying the values in the table by 2 and 3.}
\end{deluxetable}

Once these realizations are generated, we estimate the best fit parameters using our \texttt{darkmix} algorithm. As explained in section~\ref{sec:over}, this algorithm starts with an approximation of initial values, which is later optimized using the Nelder-Mead algorithm. 
The three-halos case is easy to model and the true parameters lie inside the confidence intervals of our best fit. The samples with nine-halos are more challenging, and we will present several results and plots to show the performance of our model, which we consider acceptable. With 16 halos, the model is no longer able to find and correctly model some of the extra halos because they are too small and faint to be distinguished from the background. 

\begin{figure}
\begin{center}
\includegraphics[width=.45\linewidth]{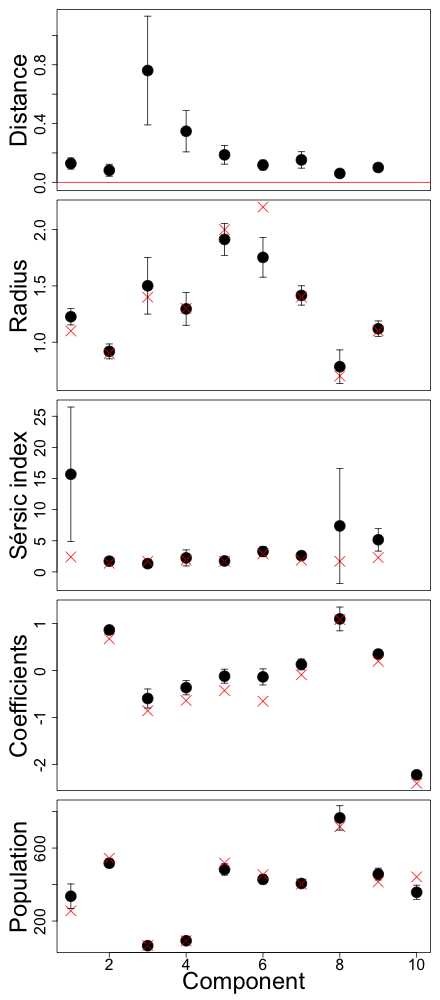}
\includegraphics[width=.45\linewidth]{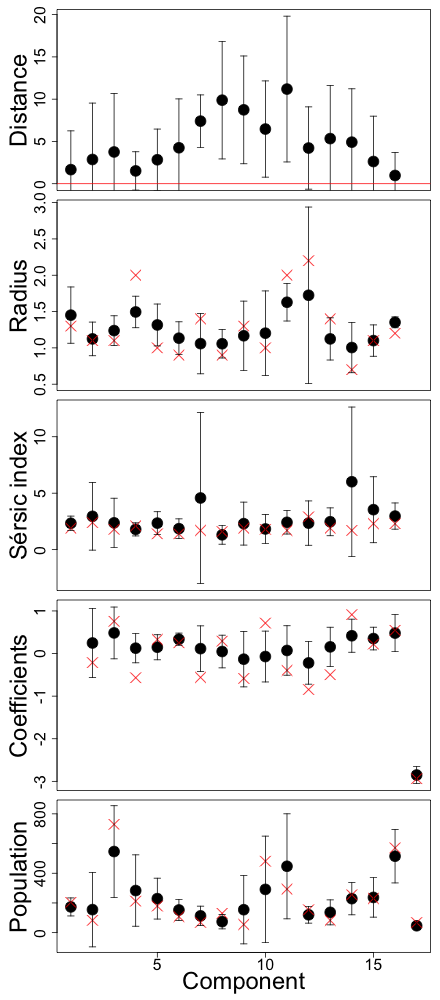}
\end{center}
\caption{Left: distribution of best fit parameters for the 20 realizations of 9 halos with low density (black) and true parameters from Table~\ref{tab:val9} (red crosses). The validation of the center of the halos is showed as the distance (first panel) to the true center (red line). Right: same for the 16 halos configuration (see Table~\ref{tab:val16}).}\label{plot:val}
\end{figure}

As we can see in Fig.~\ref{plot:val}, the parameters for a population of nine components is generally correctly estimated. The center of the halos is found. This allows for correct estimations of the radius $r_e$ and the Sérsic index $n$, only in two cases do we see one of these parameters to be outside the 1-sigma error bars. The mixture coefficients tend to be slightly overestimated, but a correction of these values does not greatly improves the maximum likelihood. Consequently, the estimation of the components population is rather accurate, which allows for a correct classification of the particles. Even in the case of 16 halos, when several components are incorrectly estimated, the halos' population is close to the true values.

\begin{figure}
\begin{center}
\includegraphics[width=.6\linewidth]{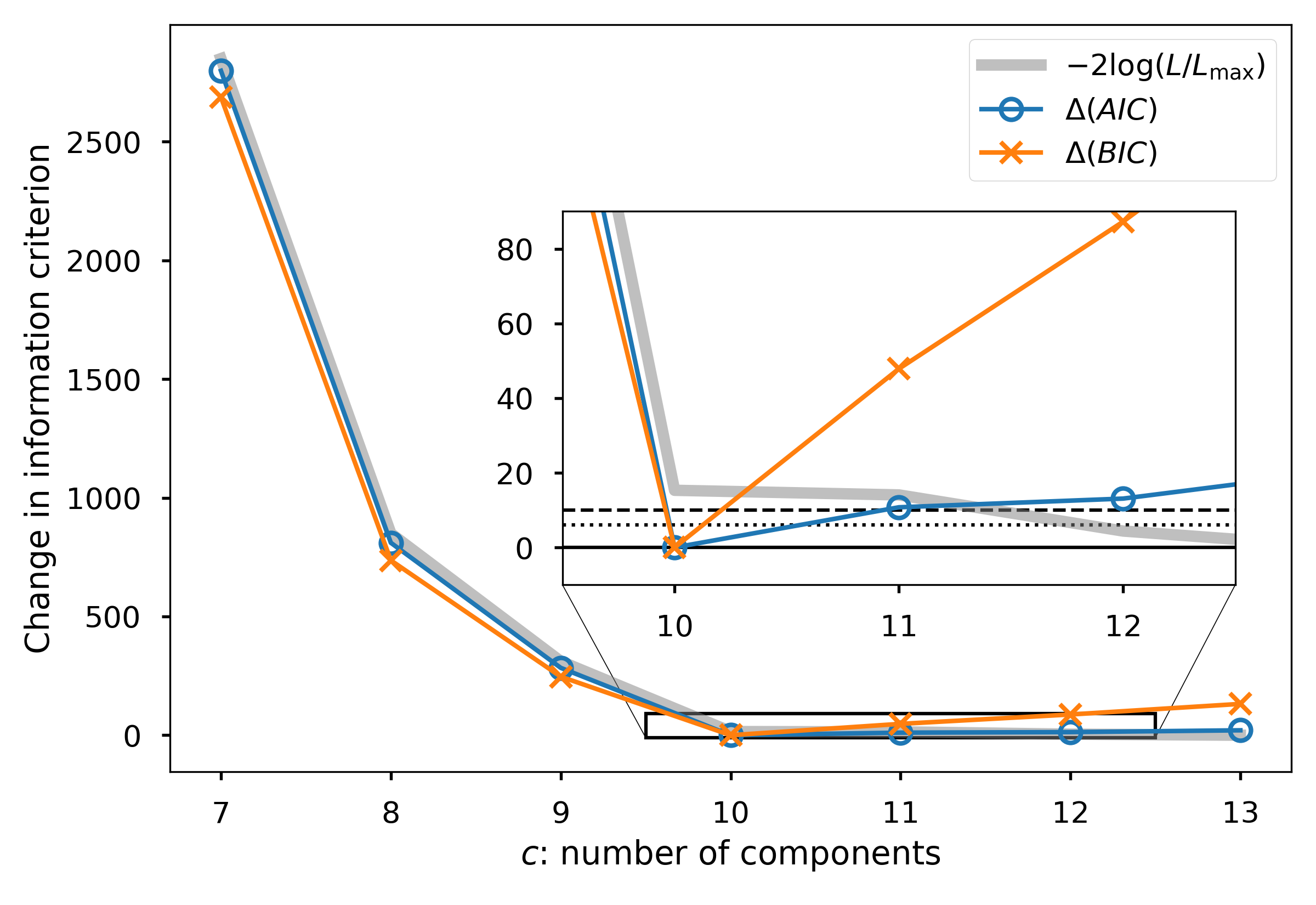}
\end{center}
\caption{Relative change in the log-likelihood (gray band), and in the information criteria BIC and AIC (symbols), as a function of the number of components included in the model, $c$. The model is fit over the nine-halo configuration.
Changes are shown with respect to the best model in each case.
The inset shows a zoom of the range $10 \leq c \leq 12$.
The horizontal dotted (dashed) line shows the threshold for strong (very strong) evidence, $\Delta (BIC) > 6$ ($> 10$) according to \citet{kass1995bayes}.
}\label{plot:aic9}
\end{figure}

An additional validation analysis was performed with the nine-halo configuration and low density: the same realization of particles (one of the 20 samples generated for the previous analysis) was estimated using mixture models of different components, from 6 to 12 halos (see Fig.~\ref{plot:aic9}). The results were shown to be satisfactory: for $k \leq 8$, the model finds the real halos and tries to estimate the Einasto profile. However, it is not until we input the right value of $k = 9$ halos that the halos are not only found but correctly estimated, as seen in Fig.~\ref{plot:val}. When $k \geq 10$, the model tries to estimate spurious over-densities in the background population, which creates halos with large radius and Sérsic index. With such values, the mass of the halo is concentrated around a few close particles, which adds a minimal contribution to the maximum likelihood of the model. This contribution is negligible, and the AIC and BIC clearly show a regression for such models.

\begin{figure}
\begin{center}
\includegraphics[width=.42\linewidth]{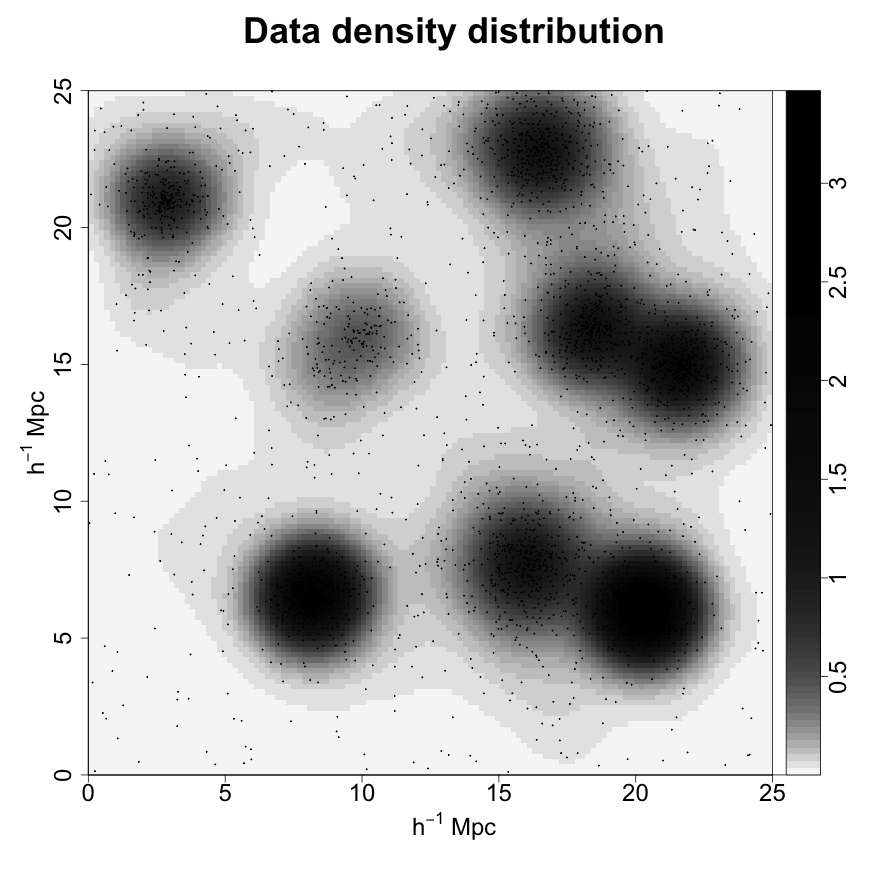}
\includegraphics[width=.42\linewidth]{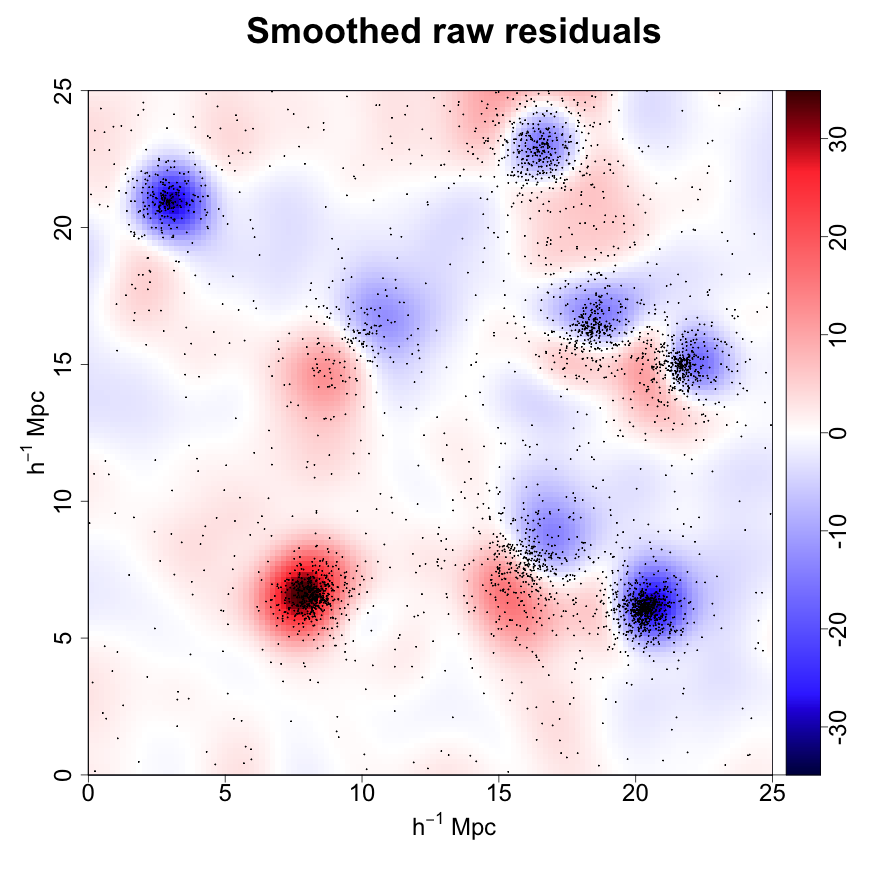} \\
\includegraphics[width=.42\linewidth]{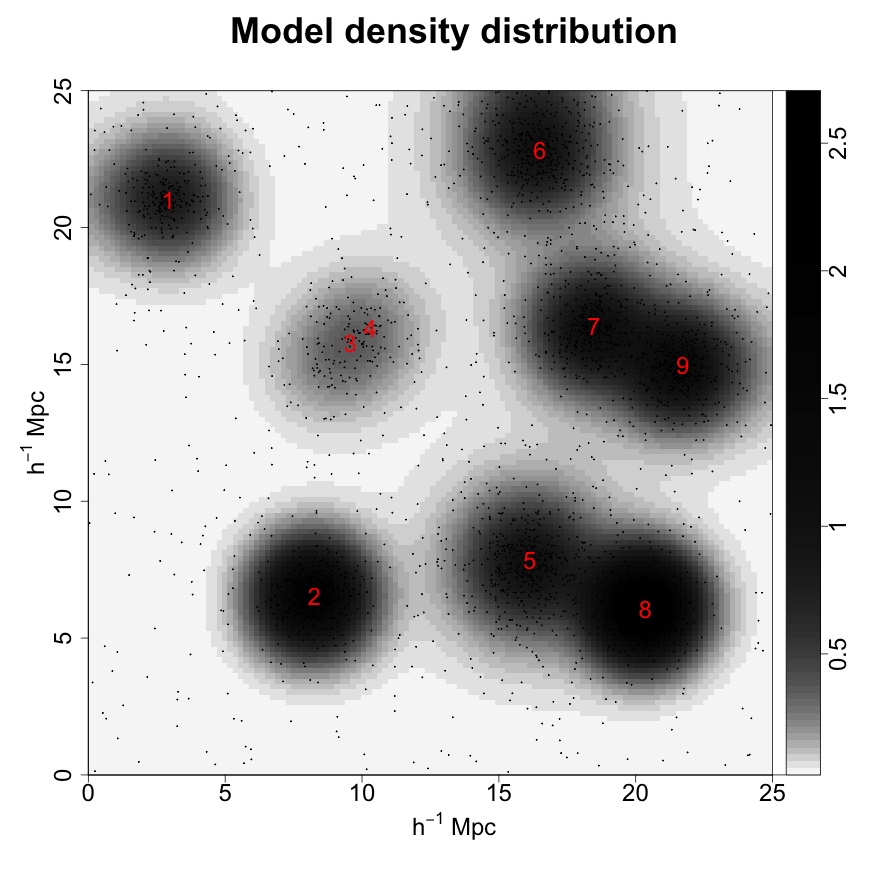}
\includegraphics[width=.42\linewidth]{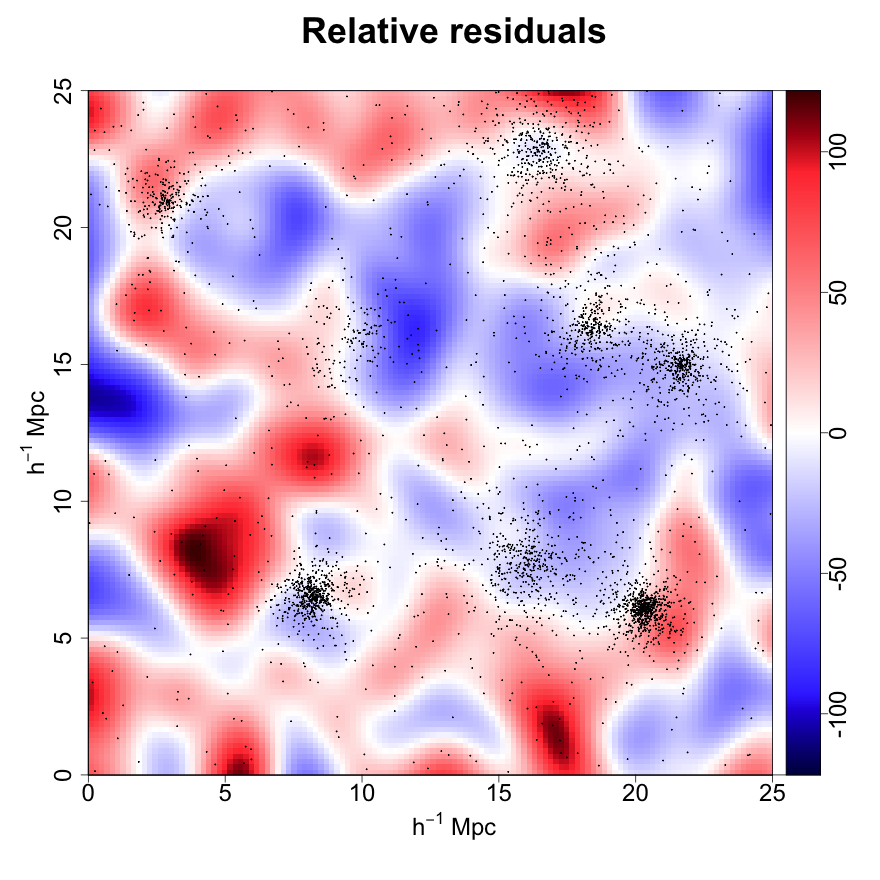}  \vspace{-0.3in}
\end{center}
\caption{Comparison of one realization of a nine-halo sample and density 0.25 with the estimated model. Top left-hand panel: data kernel particle density field $\Sigma_{\omega}^*(\mathbf{r} | \mathbf{X})$. Top right-hand panel: smoothed raw residuals $s(\mathbf{r})$. Bottom left-hand panel: model particle density field $\Sigma_{\omega}^{\dag}(\mathbf{r} | \mathbf{X}, \vec{p},\Theta)$. Bottom right-hand panel: relative residuals $e(\mathbf{r})$. In the data and model density fields, gray intensity is in logarithmic scale. Both images have been normalized over the number of particles. Dark areas indicate a denser region or higher probability of being occupied by a point, while lighter areas are unlikely to be occupied by a particle. Smoothing bandwidth for all maps is $\omega = 1$, and the values summed over the dimension Z.  Data points from Figure~\ref{md3d} are shown as small black dots.}\label{plot:maps}
\end{figure}

We conclude that our model is fully validated for configurations of at least nine halos with background component and densities equal or greater than 0.25 particles per volume unit. In Fig.~\ref{plot:maps}, we provide a four-panel plot with the data and model densities, plus the raw and relative residuals. The plot shows the close agreement between data and model. Only two halos (top left-hand a bottom left-hand panels) seem to be overestimated (blue raw residuals) and underestimated (red in raw residuals). The relative residuals show an uncorrelated pattern with the data, which means that no component is excluded from the model.

The case of 16 halos is still interesting and valid conclusions can be obtained for the richer halos, although the fainted halos might add biased results. All of the results and related data files can be found in our Git Hub repository.

\section{Data}\label{sec:data}

We perform this analysis for a case study data set from the Bolshoi simulation \citep{2011ApJ...740..102K}. This cosmological N-body simulation offers the necessary conditions and a high particle resolution that allows us to apply our methodologies over a sample of scientific interest. The MultiDark Database \citep{2013AN....334..691R} hosts two 8.6 billion particle cosmological N-body simulations: the already mentioned Bolshoi simulation \citep{2011ApJ...740..102K} and MultiDark Run1 simulation (MDR1, or BigBolshoi) \citep{2012MNRAS.423.3018P}. The Bolshoi simulation can be used to study both the large scale structure of the universe and the properties of dark matter halos. In this work we focus on the latter, which agreed with the assumptions made for our model.

The MultiDark Database allows us to use a SQL (Structured Query Language) query interface to extract the desired sample. In this paper, data has been extracted from the Bolshoi simulation. The simulation has been performed in a volume of $250 \, h^{-3} \text{Mpc}^3$, having a mass resolution of $1.35 \times 10^8 \, h^{-1}$ M$_{\odot}$, and a force resolution of physical (proper) scale of  $1 \, h^{-1} \text{kpc}$ \citep{2011ApJ...740..102K}. The cosmological parameters that we have used for this simulation are $\Omega_{m} = 0.27$, $\Omega_{b} = 0.0469$, $\Omega_{\Lambda} = 0.73$, $\sigma_8 = 0.82$, spectral index $n_s = 0.95$, and $H_{0} = 100 \, h$ km $\text{s}^{-1} \text{Mpc}^{-1}$ with $h=0.70$. The snapshot that we have used is at redshift $z=0$.

Since this work is a case study, we select a small region of interest from the table \texttt{Bolshoi.Particles416} at  \texttt{https://www.cosmosim.org/}. We are interested in a volume $W$ containing an interesting structure with halos of different sizes and merging cases. With this purpose, we select a flat cuboid with an squared face to facilitate the two-dimensional examination. This sample contains three halos that are among the 100 most massive, plus several other halos of smaller size.  The final sample contains 2081 particles in a volume of $4375 \, h^{-3}$ Mpc$^3$ (defined by a box of $25 \, h^{-1}$ Mpc$\times 25 \, h^{-1}$ Mpc$ \times 7 \, h^{-1}$ Mpc). We wish to remark here that we tested the method on other more sparse galaxy samples. However, success is not reached when the data sparsity of the sample is high. Evaluating which is the minimum needed structure density to be properly described by our model is outside the scope of this work, but this can serve as a reference. The Bolshoi simulation also provides a catalog of halos that have been categorized with the BDM algorithm \citep{1997astro.ph.12217K, 2013AN....334..691R}. In section~\ref{bdm}, we make a comparison between these halos and our findings. It is worth noting here that neither the BDM catalog of halos, nor any other catalog, has been used in this work apart from at the appendix. 

\begin{figure}
\begin{center}
\includegraphics[width=.55\linewidth]{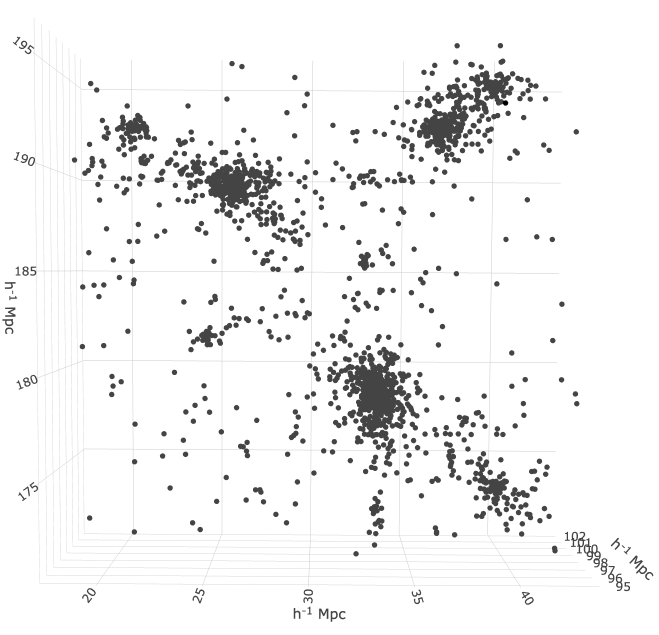}
\end{center}
\caption{Sample of dark matter particles from the Bolshoi simulation.}\label{md3d}
\end{figure}

An image of the selected sample can be seen in Figure~\ref{md3d}, notice the abundant structure and variations in the shape and size of its clusters. 

\section{DarkMix Application to the Bolshoi Simulation}
\label{sec:res}

\subsection{The Fitting Procedure}

This section serves as an example of how our code can be used to estimate a model for the data set presented in section~\ref{sec:data}. All of the calculations and results presented in this section have been obtained with \texttt{darkmix} code and a full walk-through can be found in the code documentation \footnote{https://darkmix.readthedocs.io/en/latest/darkmix\_steps.html}. 

The data can be easily loaded into a \texttt{spatsat} point process object, while the model and the parameters are defined as arrays of functions and values. The code contains functions for Einasto profiles and the background component (a constant function) but the user can easily define additional functions to similarly model a structure of interest. Regarding the integration grid, this object is used to calculate the model mass and in eq.~\ref{mmmass}. After testing different grid sizes, the authors recommend using a relation of 2:1 (i.e., we divide the sides of length 25 of our window into 50 parts). Thinner grids do not produce a different result. We only recommend using a thinner grid for plotting purposes (in this work, we use 128 grid points per side of 25 units).

Once these objects are created, we can proceed to estimate the parameters. It is advisable to start with an initial guess  of the number of halos and their centers. We recommend using function \texttt{centers}, which outputs a list with the $k$ densest locations in the data set and will be used as the initial guess for the halo components.

The remaining parameters are defaulted to $1$ for the radius $r_e$ and $3$ for the S\'ersic index $n$. The mixing coefficients are in logarithmic form: $1$ for the halos weights and $-2$ for the background component.

We will start with a $c = 11$ components model and the Nelder-Mead simplex algorithm via the R function \texttt{optim}, which maximizes the likelihood function of our mixture model. However, the model needs several iterations to achieve our best fit. To improve the estimation, additional functions have been created to freeze some parameters while estimating the rest. While algorithms spend little time finding the center of the halos, most of the computation is devoted to the estimation of the three Einasto model parameters. The radius of a component is generally independent of the center of a distant location and reducing the total number of parameters per optimization can greatly improve the fit. Each function fixed one of the different parameter types: centers, radii, S\'ersic index and mixture coefficients. It is advisable to repeat this procedure until convergence and ending by calling the \textit{optim} function with no frozen parameters for a final fit. For the data set that we have used in this work, the full procedure might take around 30 minutes to complete on a commonly-available laptop.

The results for our data set in Figure~\ref{md3d}  for an assumed $c=11$ number of components are shown in Table~\ref{dens}. The expected number of particles per component can be obtained directly from the mixture model: the component's mass is integrated independently and normalized to the total number of particles. The last column of Table \ref{dens} shows the expected number of particles per component.

Function \texttt{optim} and the other R routines that are designed to estimate the maximum likelihood provide the hessian matrix of the best fit set of parameters. This matrix can be used to obtain confidence intervals for the parameters. However, the numerical approximation of the hessian matrix obtained in our problem was ill-defined and produced negative variances. Given the impossibility of calculating the confidence intervals, we decided to run a MCMC routine with a simpler model and estimate them here. We explain this with more detail in \S\ref{simpler.sec} .

\begin{figure}
\begin{center}
\includegraphics[width=.6\linewidth]{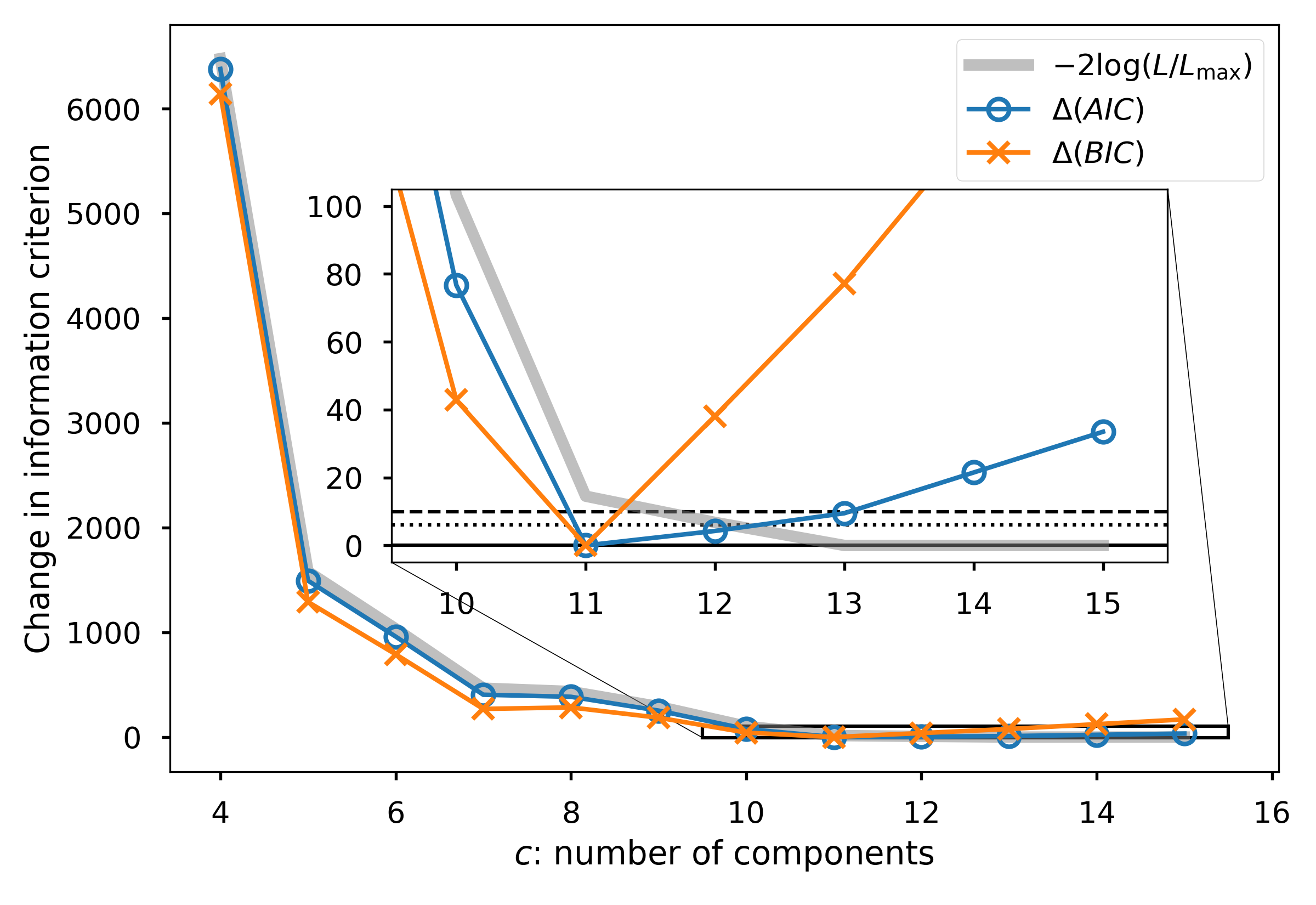}
\end{center}
\caption{Relative change in the log-likelihood (gray band), and in the information criteria BIC and AIC (symbols), as a function of the number of components included in the model, $c$. The model is fit over the Bolshoi dataset.
Changes are shown with respect to the best model in each case.
The inset shows a zoom of the range $c \geq 10$.
The horizontal dotted (dashed) line shows the threshold for strong (very strong) evidence, $\Delta (BIC) > 6$ ($> 10$) according to \citet{kass1995bayes}.
}\label{fig:aic}
\end{figure}

The mixture model should be fitted for several values of $c$, and the AIC and BIC (eq.~\ref{eq:aic} and~\ref{eq:bic}) functions should be calculated for model selection. Figure~\ref{fig:aic} shows the log-likelihood and information criteria values from fits of the Bolshoi simulation data set for models with $c=4$ to $15$ components.
As expected, the log-likelihood increases monotonically with the number of components, while both the BIC and the AIC reach a minimum value for $c=11$. Consequently, this is the preferred model according to both criteria.
We can interpret the evidence in favour of this model according to both criteria using the scale introduced by \citet{kass1995bayes}.
In the case of BIC, the $c=11$ model is very strongly favoured in all cases, as $\Delta(BIC) > 10$ for all other models.
However, when considering the AIC,  the models with $c = 12, 13$ cannot be discarded. This is a consequence of the AIC's lower penalty for increased model complexity (see eqs. \ref{eq:bic}, \ref{eq:aic}).
Following the BIC result, and choosing the most parsimonious among the accepted models by the AIC, we stick to the model with $c=11$.

As we discuss in \S\ref{simpler.sec}, it is possible for scientific considerations to prefer a non-optimal model, such as $k = 6$ rather than $k=10$.   Generally, the most populous halos will appear in all models and increasing $k$ will identify small, sparser halos.  However, it is also possible that small changes in $k$ may lead to major changes in the structure of the best-fit model.  In particular, spatial mixture models can identify a large diffuse halo that encompasses or overlaps smaller, denser halos.  This will occur when the clustering of points has a complicated hierarchical structure, rather than exhibiting distinct halo structures.  Since the model for each value of $k$ is a maximum likelihood fit, they are all statistically valid.  Whether to use the BIC and AIC for model selection or another choice of $k$ that gives a more parsimonious or more complete model of the particle distribution is a scientific decision.  

\begin{deluxetable}{lrrrrcrrrrrrr}
\caption{Maximum Likelihood Fit to the 10 Halo Model} \label{dens}
\tabletypesize{\small}
\tablehead{
& \multicolumn{4}{c}{Initial estimates} && \multicolumn{6}{c}{Best fit parameters} &\\
\cline{2-5} \cline{7-13}
\colhead{k} & \colhead{x} &\colhead{y} &\colhead{z} &\colhead{$\rho$} && \colhead{$x_0$} &\colhead{$y_0$} &\colhead{$z_0$} &\colhead{$r_e$} &\colhead{$n$} &\colhead{$\log w$} &\colhead{N}  }
\startdata
1 & $33.5$ & $178.5$ & $99.5$ & $0.078$ && $33.5$ & $178.9$ & $99.7$ & $2.1$ & $14.1$ & $0.00$ & $688$ \\
2 & $36.5$ & $192.5$ & $98.5$ & $0.060$ && $36.7$ & $192.4$ & $98.8$ & $0.9$ & $2.9$ & $0.95$ & $354$ \\
3 & $39.5$ & $174.5$ & $97.5$ & $0.023$ && $39.5$ & $174.3$ & $97.5$ & $1.7$ & $28.5$ & $-0.37$ & $138$ \\
4 & $25.5$ & $189.5$ & $98.5$ & $0.018$ && $26.0$ & $189.4$ & $98.9$ & $2.0$ & $26.2$ & $-0.14$ & $449$ \\
5 & $20.5$ & $192.5$ &$100.5$ & $0.015$ && $20.4$ & $192.8$ & $100.9$ & $1.6$ & $28.3$ & $-0.53$ & $94$ \\
6 & $38.5$ & $193.5$ & $96.5$ & $0.011$ && $38.7$ & $193.9$ & $96.3$ & $1.4$ & $4.0$ & $-0.06$ & $119$ \\
7 & $37.5$ & $175.5$ & $99.5$ & $0.005$ && $37.4$ & $175.4$ & $99.5$ & $8.7$ & $9.2$ & $-2.75$ & $47$ \\
8 & $33.5$ & $172.5$ & $100.5$ & $0.005$ && $33.1$ & $190.4$ & $99.4$ & $2.0$ & $26.9$ & $-1.43$ & $23$ \\
9 & $32.5$ & $185.5$ & $99.5$ & $0.005$ && $32.8$ & $185.5$ & $100.0$ & $3.8$ & $30.0$ & $-1.92$ & $48$ \\
10& $21.5$ & $190.5$ & $98.5$ & $0.004$ && $25.1$ & $181.9$ & $95.3$ & $4.1$ & $25.3$ & $-1.91$ & $48$ \\
Bk & \nodata & \nodata & \nodata & \nodata && \nodata & \nodata & \nodata & \nodata & \nodata & $0.13$ & $73$ 
\enddata
\tablecomments{Input point process shown in Figure~\ref{md3d}. $k$ identifies the halo component.  Initial $\mathbf{r_0}$ values from kernel density estimator in order of decreasing maximum density $\rho$. Best fit parameters give the halo center, Einasto parameters $r_e$ and $n$, mixing coefficient, and number of dark matter particles.}
\end{deluxetable}  
 
\subsection{Particle Membership in Halos}\label{sec:mem}

Once we have our estimated mixture model, we may want to classify the data set particles into its different components. This comes naturally with a soft classifier method such as a mixture model: given a particle, we evaluate each model component at the particle location and we then normalize the obtained quantities to one. The resulting values can be understood as membership probabilities and used in a multinomial distribution to assign one particle to $c$ components with the $c$ probabilities. If the model is correct, then the number of particles assigned per component should match, on average, to that shown in Table~\ref{dens}.

We recommend adapting this criteria to our scientific interests. Halos with heavy tails might populate areas with particles far from the real halo boundaries, hence overestimating the halo population. In contrast, under merging circumstances, several halos might be competing for the same particle and the model will assign low probabilities to each of them. Assuming same size halos, the border between two merging halos will be populated by particles with 0.5 probability of belonging to each halos, 0.33 for three halos, etc. The multinomial criteria might end up assigning these particles to the background instead, especially when the number of merging halos is high. To compensate for this, we provide two modifications to the multinomial criteria. First, the background component will not compete with the halos and  the particle will be directly assigned to the background whenever the background probability is higher than that of any halo. Second, the user can input a threshold value such that  the particle is assigned directly to the background component if all probabilities are below this value. 

In our case study data set, we detect cases of the merging halos (see Figure~\ref{mdpg}). Following the rule given above, we recommend a threshold of 0.3, which is sufficient to assign to the background all particles in an undecided situation.

In Table~\ref{memb.tbl}, we provide an example of this classification procedure for the first five particles in our data set, showing the probability of each particle (rows) to belonging to each component (column). A final column gives the identifying number of the halo component for each particle according to the chosen decision rule. This table is used to plot Figure~\ref{mdpg}, with each component given in a different color.

\begin{deluxetable}{crrrrrrrrrrrrrrrc}
\caption{Halo Membership Probabilities for Bolshoi Dark Matter Particles} \label{memb.tbl}
\tablehead{
\colhead{Part} & \multicolumn{3}{c}{Location} && \multicolumn{10}{c}{Halo Component} & \colhead{Bkgd} & \colhead{Memb} \\  \cline{2-4} \cline{6-15} 
& \colhead{x} & \colhead{y} & \colhead{z} &&\colhead{1} & \colhead{2} & \colhead{3} & \colhead{4} & \colhead{5} & \colhead{6} & \colhead{7} & 
\colhead{8} & \colhead{9} & \colhead{10} &&   }
\startdata
1 & 39.125 & 192.997 & 95.635 && 0.001 & 0.020 & 0.000 & 0.001 & 0.000 & 0.969 & 0.000 & 0.000 & 0.000 & 0.000 & 0.007 & 6 \\
2 & 35.139 & 191.504 & 98.906 &&  0 004 & 0.942 & 0.000 & 0.007 & 0.000 & 0.010 & 0.000 & 0.022 & 0.003 & 0.000 & 0.010 & 2 \\
3 & 34.817 & 191.624 & 98.213 &&  0.006 & 0.896 & 0.000 & 0.013 & 0.001 & 0.021 & 0.000 & 0.038 & 0.004 & 0.002 & 0.018 & 2 \\
4 & 32.199 & 189.673 & 99.982 &&  0.028 & 0.029 & 0.001 & 0.112 & 0.003 & 0.004 & 0.002 & 0.730 & 0.044 & 0.002 & 0.042 & 8 \\
5 & 31.930 & 189.982 & 99.307d &&  0.024 & 0.029 & 0.001 & 0.127 & 0.003 & 0.005 & 0.001 & 0.732 & 0.032 & 0.001 & 0.021 & 8 \\
\enddata
\tablecomments{This only shows a portion of the membership table.  The full table is available in \url{https://github.com/LluisHGil/darkmix/blob/master/Output/membership.txt}.}
\end{deluxetable}

\subsection{Goodness-of-Fit and Residual Analysis}

\begin{figure}
\begin{center}
\includegraphics[width=.5\linewidth]{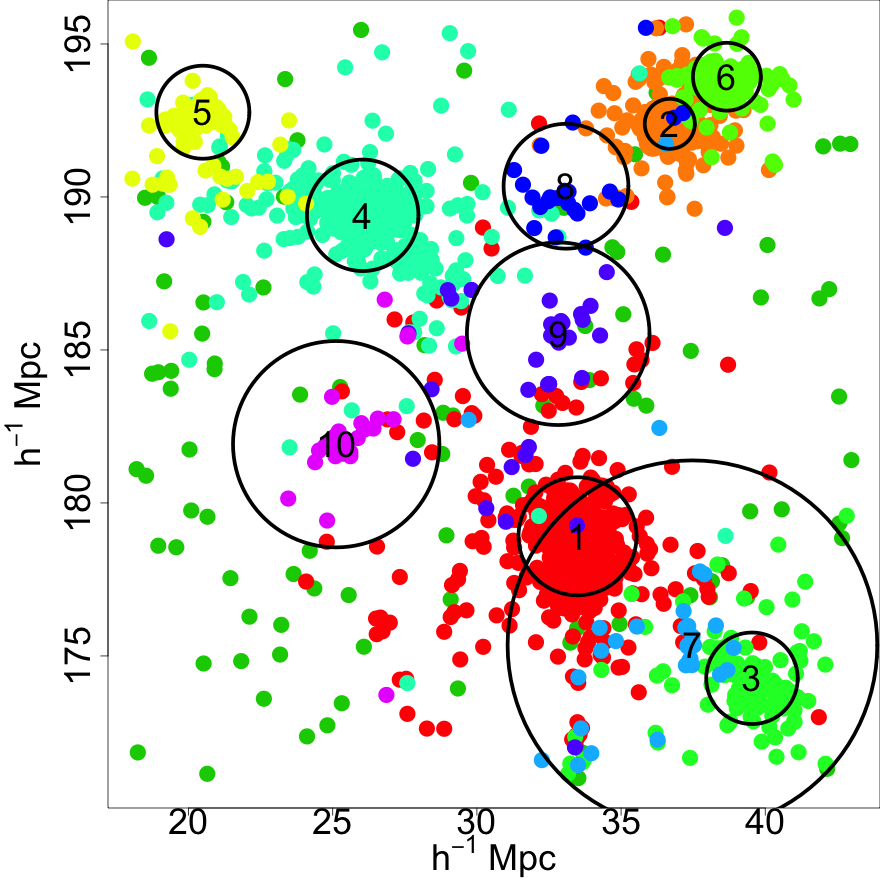}
\end{center}
\caption{Particle classification based on Table~\ref{memb.tbl}. Clusters are labeled as in Table~\ref{dens} and the circle radii are the fitted Einasto radii $r_e$.}\label{mdpg} 
\end{figure}

Once our best fit parameters are estimated, they can be displayed and analyzed for astrophysical properties. Some halo components show degenerate values for $r_e$ and $n$. As can be seen in Figure~\ref{mdpg}, each component is numbered according to Table~\ref{dens} and the circles have radius $r_e$. Halos 7 to 10 have extremely large radius. From Table \ref{dens}, we also note that these components have mixing coefficients $>20$ times lower than the other components---they have low central densities with large radii and few particles. This is a sign that these components do not follow an Einasto profile and cannot be correctly modeled by our model. In addition, the S\'ersic index $n$ is also unreliable---while values around 3 are expected in astrophysics \citep{2006AJ....132.2685M}, much greater values are found instead. In \S\ref{simpler.sec}, we provide confidence intervals for the parameters and a covariance matrix can be used to evaluate the reliability of the coefficients.

An Einasto profile with a high $n$ mimics a power law with a strong concentration of points around the center and weak tails. Optimally, we would use a different function to model this kind of structure, but we can choose between including degenerated halos or neglecting them and incorporate their particles to the background.

For our best fit model of 11 components, we have a coefficient of determination $R^2 = 0.92$. In this calculation, the kernel density estimator of the data (eq.~\ref{ker}) is compared to the model density field (eq.~\ref{model}). Even if these fields are expressed in different units, they must follow a linear relation if the fitting is perfect. A linear regression between the densities is fitted in each spatial tile and $R^2$ is reported as a measure of goodness-of-fit. 

\begin{figure}
\begin{center}
\includegraphics[width=.42\linewidth]{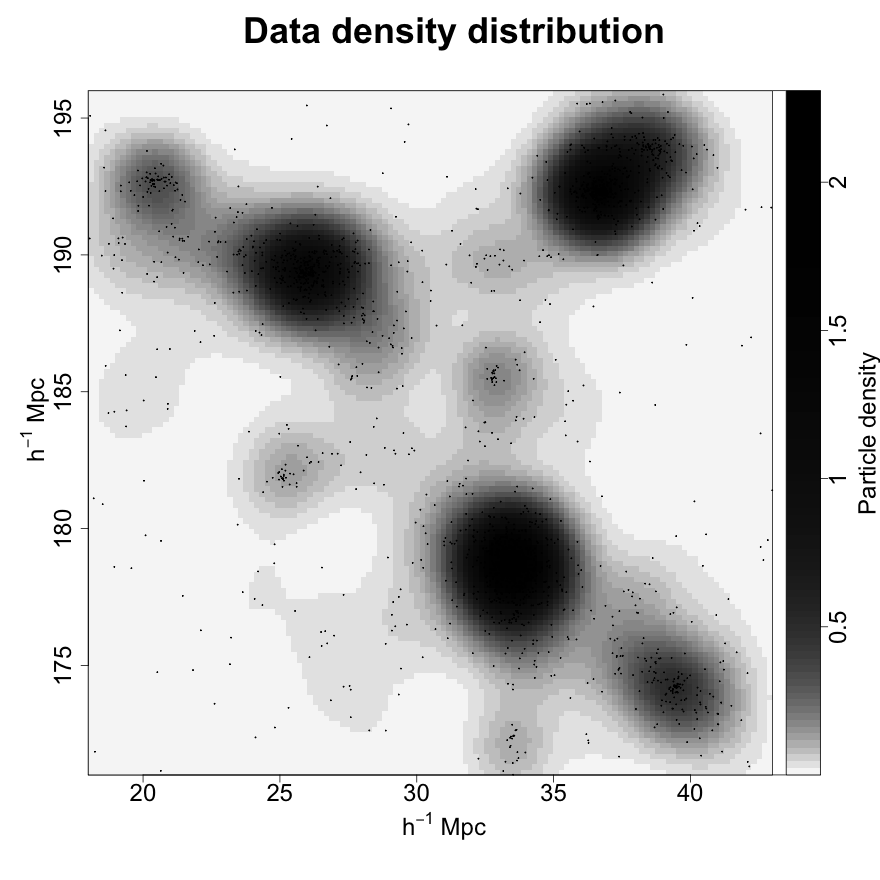}
\includegraphics[width=.42\linewidth]{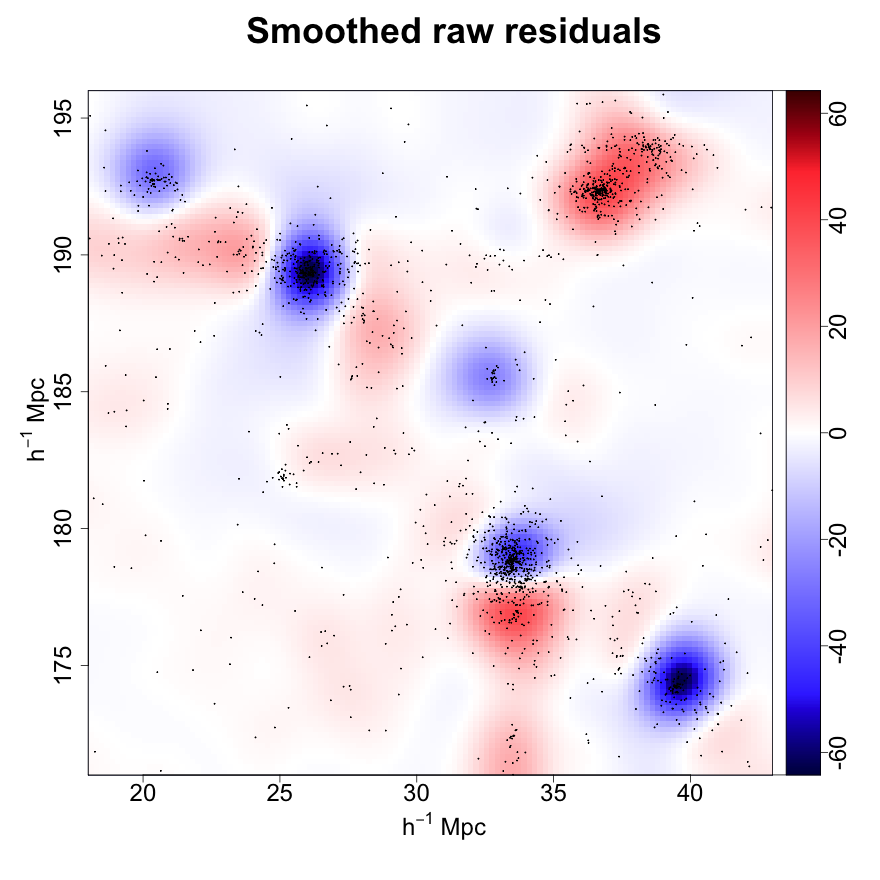} \\
\includegraphics[width=.42\linewidth]{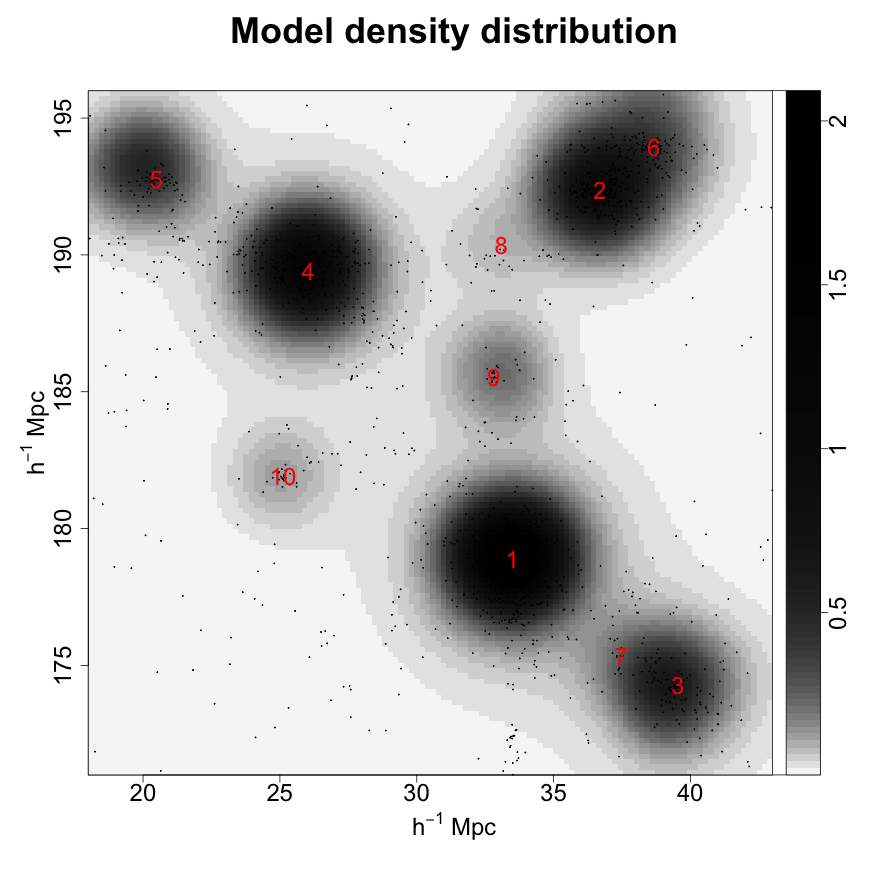}
\includegraphics[width=.42\linewidth]{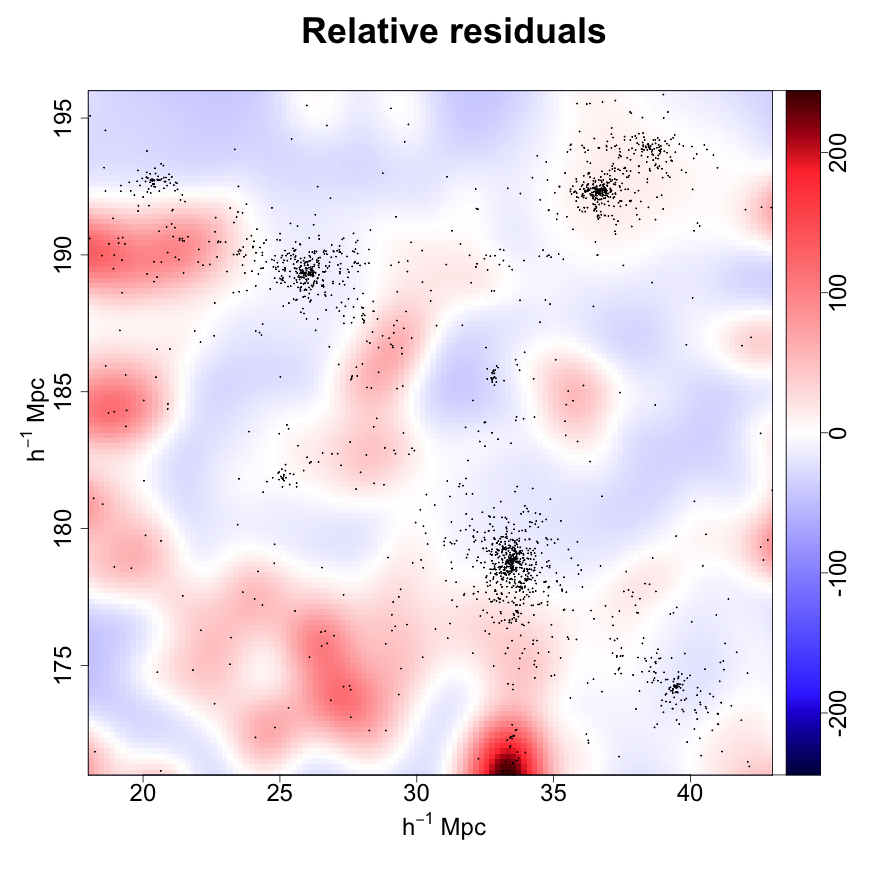}  \vspace{-0.3in}
\end{center}
\caption{Comparison of smoothed Bolshoi sample data and the 10-halo model.  Top left-hand panel: data kernel particle} density field $\Sigma_{\omega}^*(\mathbf{r} | \mathbf{X})$. Top right-hand panel: smoothed raw residuals $s(\mathbf{r})$. Bottom left-hand panel: model particle density field $\Sigma_{\omega}^{\dag}(\mathbf{r} | \mathbf{X}, \vec{p},\Theta)$. Bottom right-hand panel: relative residuals $e(\mathbf{r})$. In the data and model density fields, gray intensity is in logarithmic scale. Both images have been normalized over the number of particles. Dark areas indicate a denser region or higher probability of being occupied by a point, while lighter areas are unlikely of being occupied by a particle. The smoothing bandwidth for all maps is $\omega = 1$, and the values summed over the dimension Z.  Data points from Figure~\ref{md3d} are shown as small black dots.\label{data_model_dens}
\end{figure}

These two fields, $\Sigma_{\omega}^*$ and $\Sigma_{\omega}^{\dag}$, can be seen in the left-hand panel of Fig.~\ref{data_model_dens}. We can see a clear agreement in both the distribution, size and range of densities of the halos. The main differences arise from asymmetries in the real structure, which are difficult to capture with spherical halos.

\begin{figure}
\begin{center}
\includegraphics[width=.52\linewidth]{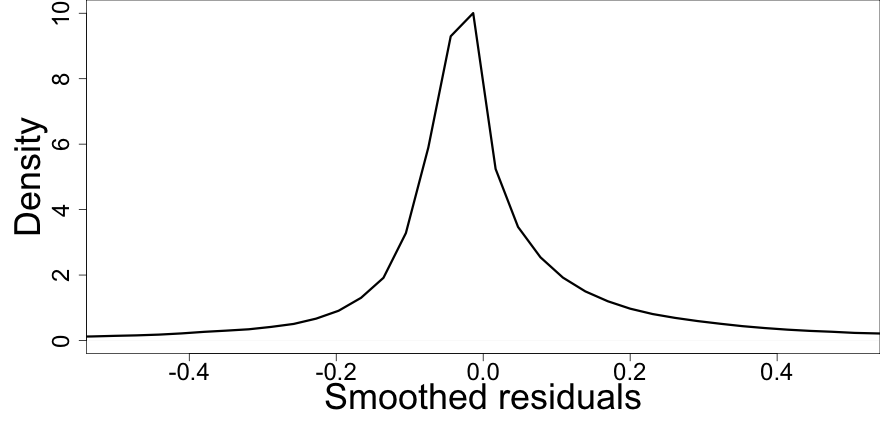}
\end{center}
\caption{Probability density distribution of the smoothed raw residuals.}\label{res_dens}
\end{figure}

We can see the probability density distribution of the smoothed raw residuals in Fig.~\ref{res_dens}. The residuals are skewed ($g_1 = -5.4$) at low values away from the halo peaks and exhibit heavy tails at high values near halo peaks. The reason for the heavy tails is evident in the lower left-hand panel, which shows the absolute ($s(\mathbf{r})$) from equations~\ref{eq:su}. While errors have mean zero by construction, a strong red-blue pattern around one of the biggest halos is responsible for the largest residuals. This suggests that the shape is poorly fitted by an Einasto profile. It is difficult to assess if these errors are within the expected shot-noise errors of the model or if we should consider a change in the model, such as a parameter fit refining or a different profile. However, the bottom right-hand panel of Fig.~\ref{data_model_dens}, which maps the relative raw residuals ($e(\mathbf{r})$), shows that these model errors are small compared to the background, which is covered by red areas when data is present. This implies that, in relative terms, the background has stronger fitting problems that the halo components.

This panel is particularly effective at revealing structures that are not included in the model. The most prominent structure missing in the model is at the bottom where, due to truncation by the window edge, an Einasto profile of a sparse cluster did not fit well. Finally, on Fig.~\ref{mdpg} we present a randomized classification of the particles based on our model. The plot shows how a soft classifier mixes the particles of different components, specially under merging conditions.

\begin{figure}
\begin{center}
\includegraphics[width=.42\linewidth]{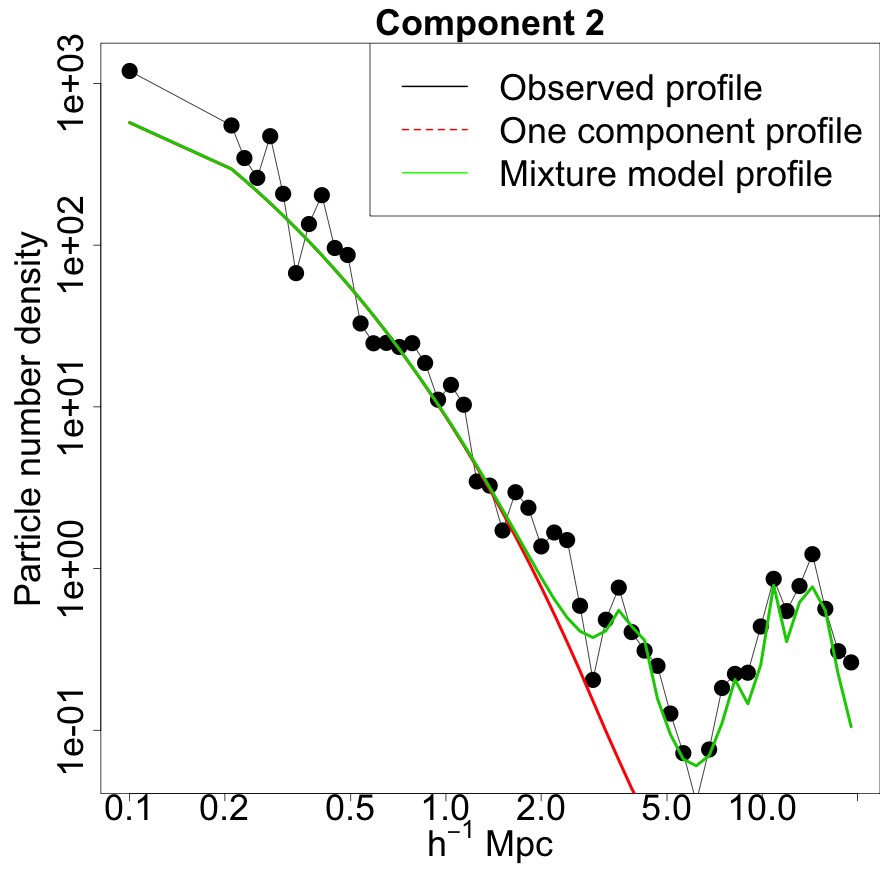}
\includegraphics[width=.42\linewidth]{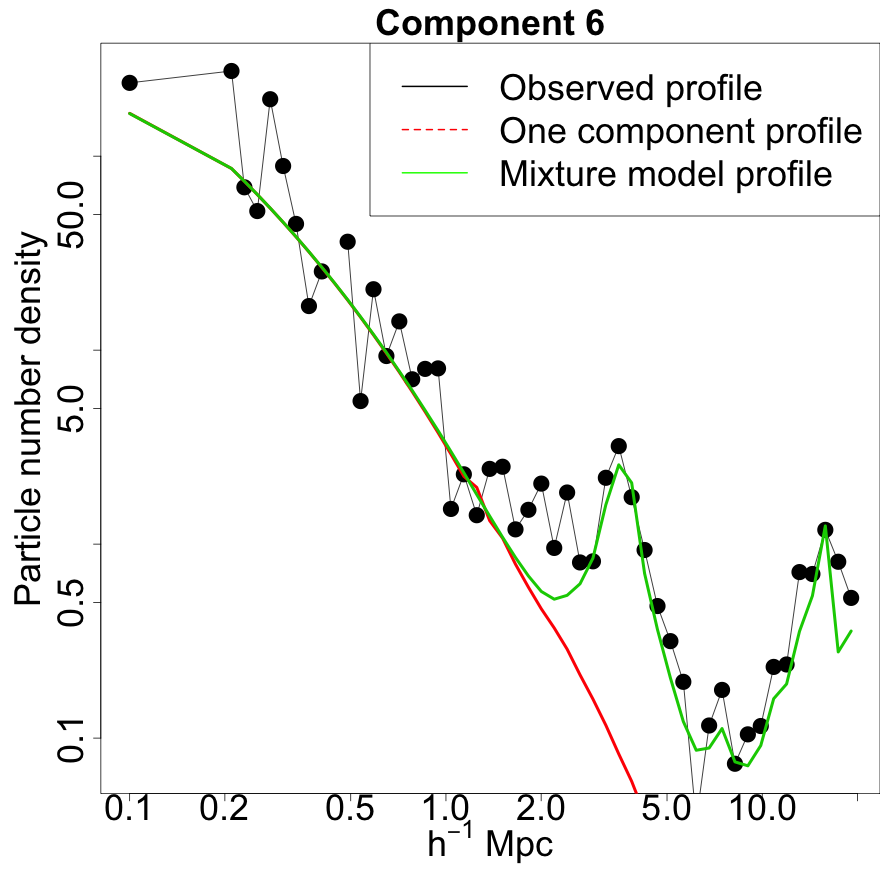}
\end{center}
\caption{Radial density profiles of clusters 2 (left-hand panel) and 6 (right-hand panel). Black dots represent the particle number densities. The red curve is the density profile of an isolated component ($\hat{\rho}(r | \mathbf{r_{0,j}})$) and the green curve is the density profile of the full mixture model ($\hat{P}(r | \mathbf{r_{0,j}})$). Note the close fitting of the full mixture model profile (in green).}\label{mdpc}
\end{figure}

Another interesting visualization analysis is the profile extraction. With eq.~\ref{pro:fit} and~\ref{pro:com} we can plot the one-dimensional profile of the whole mixture model and also that of any individual halo component. In Fig.~\ref{mdpc}, we see the case for components 2 and 4 in our classification. We can see that the Einasto profiles (red curves) match the empirical profile (black dots) out to $\sim 2$~Mpc, beyond which additional halos appear. The full model (green curves) follows the empirical profile remarkably well, with peaks associated with other clusters. With this method, mixture models can be used to disentangle the real profile of merging halos, recovering the distribution of these structures where the empirical profile does not allow it.

\subsection{Comparison with BDM Calculations}\label{bdm}

The tables provided by the MultiDark Database \citep{2013AN....334..691R}, including the Bolshoi simulation used in this work, include a catalog of halos that have been detected with the Bound Density Maximum (BDM) algorithm \citep{1997astro.ph.12217K, 2013AN....334..691R}. This algorithm detects local density maxima and defines a spherical halo that removes unbound particles. This algorithm allows for the detection of subhalos, which are smaller structures inside parent halos.

\begin{figure}
\begin{center}
\includegraphics[width=.47\linewidth]{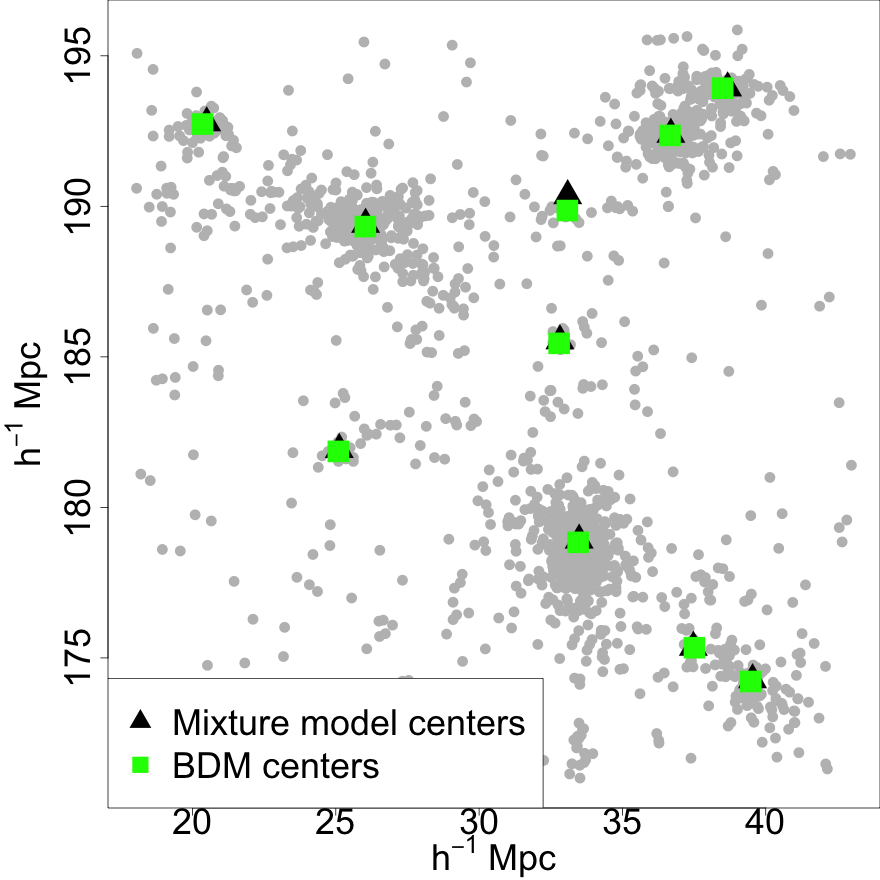}
\includegraphics[width=.47\linewidth]{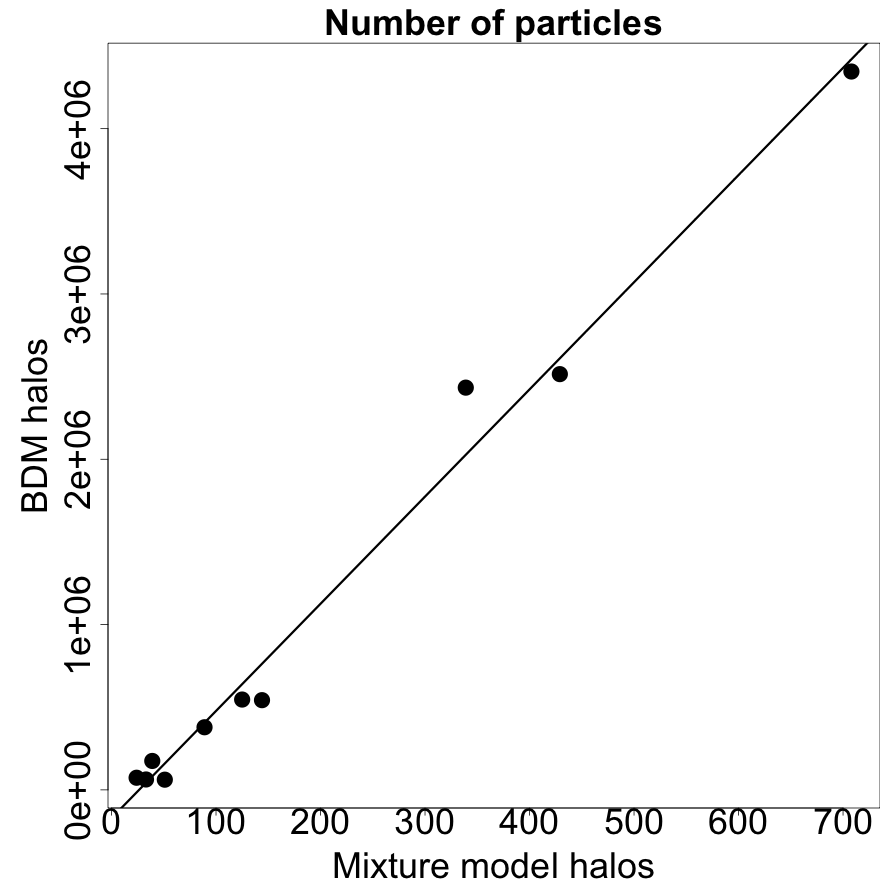}
\end{center}
\caption{Left-hand panel: layout of dark matter particles with BDM halo centers (squares) and our estimations (triangles). Right-hand panel: number of particles found in the BDM catalog vs estimated number of particles in our mixture model. Line indicates the best linear regression between both variables.}\label{bdm_plot}
\end{figure}

Our work makes use of a low density data sample, which does not permit the detection of small structures, such as the less massive halos. Therefore, we compare our detected halos with the most massive BDM structures. In table \texttt{Bolshoi.BDMV} from the Multidark Database, and inside the region of our data set, 26 halos can be found with more than 30,000 particles. Since we have 10 halo components in our best fit model, we select the 10 halos laying closer to our centers. The resulting list can be found in the Github repository, file \texttt{bdm\_halo.txt}. In Figure~\ref{bdm_plot} we compare both sets of halos. As can be  seen, the selected BDM halos lie very close to our halo centers and have been calculated to contain a number of particles that highly correlates with our estimation (see Table~\ref{dens}, column $N$). Since we are using just a sample of particles to illustrate the methods, our numbers are only a fraction of the real number of particles.

\subsection{A Simpler, Bayesian Model} \label{simpler.sec}

In Fig.~\ref{fig:aic} we see how the AIC and BIC showed a plateau after $c=6$, Fig.~\ref{mdpg} shows how the 7$^{th}$ component has a radius $r_e$ that extends around other halos and does not follow an Einasto profile, and components $8-10$ are sparse.  It may thus be scientifically useful to examine a simpler model with 7 rather than 11 components. 

This simplified model still has 36 parameters and was estimated using the MCMC sampling approach with maximum a posterior best fit parameters, where 3,600,000 iterations are needed to satisfactorily map the distributions. Although it is not included in darkmix.R and shown here, we recommend the user to examine graphical and scalar (e.g., the Gelman-Rubin statistic) diagnostics of convergence of the MCMC chains. CRAN package \textit{CODA} is widely used for this purpose. The long calculation time can be a considerable handicap and the MLE procedure that was given earlier may be operationally more feasible for large data sets.   The advantage of a Bayesian approach is to map non-Gaussianities in the posterior distributions and curved relationships in bivariate parameter confidence intervals. However, for many science applications these capabilities are not needed and the MLE approach would be preferred.  

\begin{deluxetable}{crrrrrrr}
\tablecaption{Bayesian Fit to a Simpler c=7 Model}  \label{tab:6k}
\tablehead{
\colhead{$k$} & \colhead{$x_0$} & \colhead{$y_0$} & \colhead{$z_0$} & \colhead{$r_e$} & \colhead{$n$} & \colhead{$\log w$} & \colhead{$N$} }
\startdata
1 & $33.46 \pm 0.01$ & $178.80 \pm 0.02$ & $99.70 \pm 0.02$ & $1.05 \pm 0.04$ & $2.4 \pm 0.2$ & $1.5 \pm 0.9$ ~~~~ & $599 \pm 22$\\
2 & $36.67 \pm 0.02$ & $192.33 \pm 0.01$ & $98.75 \pm 0.02$ & $0.77 \pm 0.05$ & $3.0 \pm 0.3$ & $2.0 \pm 1.0$ ~~~~ & $413 \pm 19$\\
3 & $26.00 \pm 0.03$ & $189.35 \pm 0.03$ & $98.87 \pm 0.02$ & $1.11 \pm 0.07$ & $2.5 \pm 0.3$ & $0.8 \pm 0.4$ ~~~~ & $376 \pm 19$\\
4 & $38.60 \pm 0.04$ & $193.88 \pm 0.03$ & $96.18 \pm 0.03$ & $0.75 \pm 0.08$ & $2.2 \pm 0.4$ & $0.7 \pm 0.5$ ~~~~ & $108 \pm 11$\\
5 & $39.47 \pm 0.02$ & $174.22 \pm 0.02$ & $97.44 \pm 0.03$ & $0.90 \pm 0.10$ & $3.2 \pm 0.6$ & $0.4 \pm 0.3$ ~~~~& $127 \pm 12$\\
6 & $20.40 \pm 0.02$ & $192.72 \pm 0.02$ &$100.81\pm 0.02$ & $0.56 \pm 0.08$ & $3.2 \pm 0.9$ & $1.0$~~~~~~~~~&   $74 \pm 8$\\
Bk & \nodata & \nodata & \nodata & \nodata & \nodata & $0.14 \pm 0.004$ & $384 \pm 26$ \\
\enddata
\tablecomments{Values are the mean and standard variation of the posterior distribution.  Component 6 has been used as the first component in the model, and therefore $w_6 = 1$. }
\end{deluxetable}

\begin{figure}
\begin{center}
\includegraphics[width=.42\linewidth]{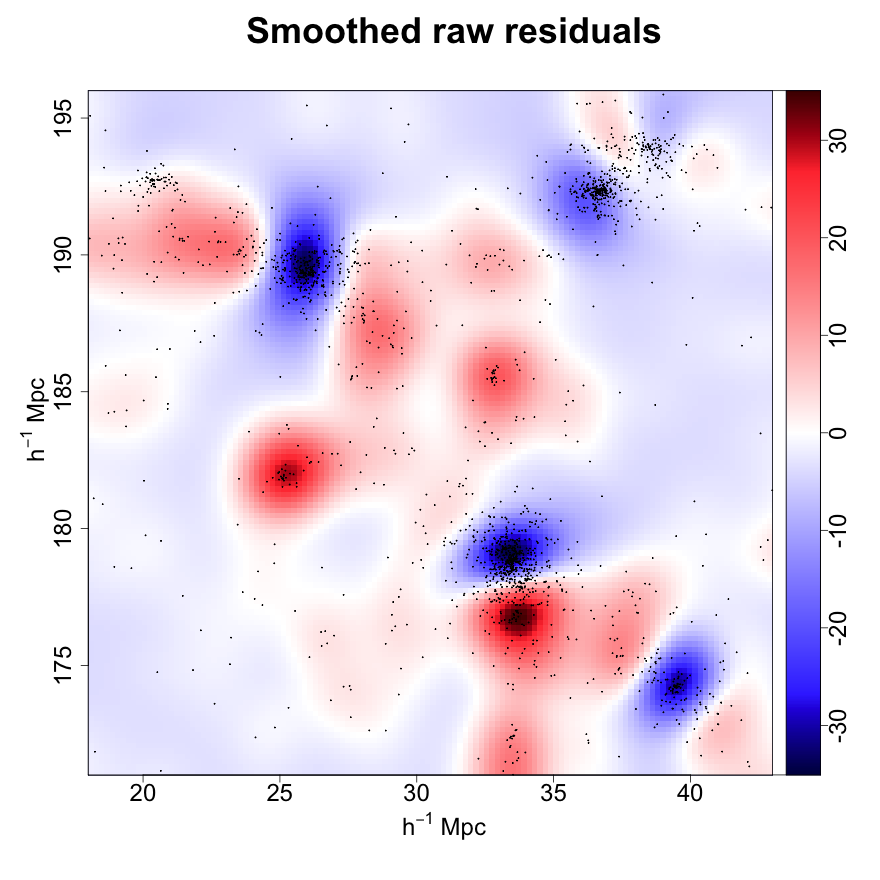}
\includegraphics[width=.42\linewidth]{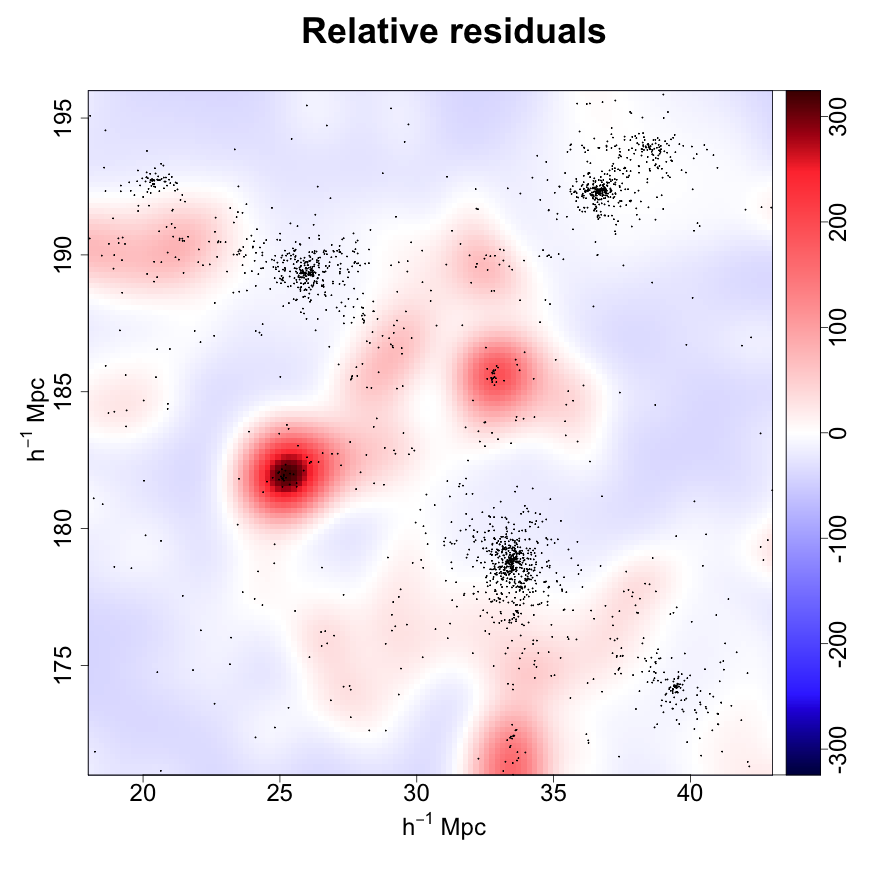}
\end{center}
\caption{Residuals from the Bayesian fitting of the $k=6$ halo mixture model.  Absolute raw residuals $s(\mathbf{r})$ (left-hand panel) and relative residuals $e(\mathbf{r})$ (right-hand panel) of the Bolshoi sample. Smoothing bandwidth is $\omega = 1$ and values summed over dimension Z.}\label{fig:6k}
\end{figure}

In Table~\ref{tab:6k} we summarize the estimated maximum a posterior estimated parameters with error bars.

Compared to the $c=11$ mixture model, the removal of components 7 to 10 has an impact on the estimation of the other parameters. The $R^2$ coefficient is  0.936, which is similar to the more elaborate model.  We show in Figure~\ref{fig:6k} the absolute and the relative residuals ($s(\mathbf{r})$ and $e(\mathbf{r})$). As can be seen, the structures that were previously modeled by components 7 and 8 appear now in red colors. The absence of model components leave these structures underestimated. However, its intensity is similar to that of some other regions in the data set and a visual inspection alone might not be enough to detect them. We use the relative residuals in Figure~\ref{fig:6k} bottom panel to detect their presence. 

\section{Conclusions}\label{sec:con}

We have demonstrated the use of finite mixture models to detect and characterize a sample of simulated dark matter halos.  The maximum-likelihood solution to a parametric model of the particle distribution produces a maximum likelihood probability density function.  Two results emerge from the model: a list of halos with their properties (Table~\ref{dens}), and the probability that each particle is a member of each cluster (Table~\ref{memb.tbl}).  This is a `soft classification' measure and additional decision rules are needed to assign membership probabilities to each particle. A variety of graphical and statistical measures of goodness-of-fit  are provided.  This mixture model approach has several important differences from other conventional clustering techniques.
\begin{enumerate}

\item The user specifies a radial profile function based on previous astronomical experience or astrophysical insight.  Here, we chose an Einasto profile, which is the three-dimensional analog of the S\'ersic profile and a generalization of the de Vaucouleurs profile for elliptical galaxies. This permits an estimation of parameters with a physical interpretation, such as the size, shape and particle abundance of a halo. In addition, this estimation is done directly on the three-dimensional data set, without the need to integrate the one-dimensional profile and the loss of information that it involves. 

\item No threshold densities, maximum distance scales, minimum membership population, or other arbitrary parameters are needed, as in nonparametric clustering procedures such as the `friends-of-friends' algorithm (single-linkage hierarchical clustering) or DBSCAN. A unique MLE for each value of $k$ is calculated without any free parameters. Model complexity (i.e., the value of $k$) is chosen based on penalized likelihood information criteria combined with scientific judgment. The location, size and population of each halo is not subject to arbitrary parameter choices but are outcomes of a likelihood calculation based on an astrophysically reasonable profile function. Only the membership of each particle can be enhanced using a threshold determined by the number of merging halos. When two or three halos overlap, a value of 0.3 is advised. 

\item As a soft classification method, we have a measure of the reliability that a given dark matter particle is a member of a given halo.  Researchers who are interested in merging scenarios might use these results in various ways to obtain more informative descriptions of the interactions. For example, sparse satellite halos can be identified and their merging history into large protogalaxies can be traced. The ability of the mixture model to identify clustered ensembles of halos (see Figure~\ref{mdpg}) may be useful. Mixture modeling of time sequences of the dark matter distribution can reveal, in a statistically rigorous fashion, when and where merging occurs.  Alternatively, decision rules can be constructed to identify the most isolated dark matter particles to understand the evolution of particles that have not merged into equilibrated halos.  

\item The parametric mixture model would be challenged in situations where halos with a wide range of populations $N$ are present. If, for example, the data set here was dominated by a large halo with $N \sim 10,000$ particles, then the likelihood may not be sufficiently improved by the addition of a small halo with $N \sim 100$.  However, most nonparametric clustering techniques have similar difficulties.  Methods such as hierarchical density-based clustering seek to address this difficulty \citep{Campello13}. 
\end{enumerate}

Mixture models have been discouraged for use to understand the galaxy large-scale structure distribution because it is dominated by interconnected curved filamentary structures rather than centrally condensed clusters \citep{KuhnFeigelson19}.  However, the dark matter collapses into distinct equilibrated halos early in the evolution of cosmic structures and is more amenable to mixture analysis. The biggest handicap of these models is the calculation costs for large data sets and large $k$.  This method is best applied to the understanding of dark matter evolution within small boxes, rather than analysis of a full multi-billion particle simulation.  There is also the possibility that convergence will be difficult, either for the optimization of the MLE or for the MCMC chains of the Bayesian calculation.  Other algorithms, such as MultiNest \citep{skilling04, ferozskilling13} or PolyChord \citep{handley15a,handley15b}, that are specially meant to estimate high-dimensional multimodal likelihood functions might be tried.  Finally, we reiterate that all of the R functions used in this work can be downloaded from our GitHub repository \texttt{https://github.com/LluisHGil/darkmix}.

\section*{Acknowledgements}
This work has been funded by the project PID2019-109592GB-I00/AEI/10.13039/501100011033 from the Spanish Ministerio de Ciencia e Innovaci\'on - Agencia Estatal de Investigaci\'on, by the Project of excellence Prometeo/2020/085 from the Conselleria d'Innovaci\'o, Universitats, Ci\`encia i Societat Digital de la Generalitat Valenciana, and by the Acci\'on Especial UV-INV-AE19-1199364 from the Vicerrectorado de Investigaci\'on de la Universitat de Val\`encia.

The CosmoSim database used in this paper is a service by the Leibniz-Institute for Astrophysics Potsdam (AIP).  The MultiDark database was developed in cooperation with the Spanish MultiDark Consolider Project CSD2009-00064.  The Bolshoi and MultiDark simulations have been performed within the Bolshoi project of the University of California High-Performance AstroComputing Center (UC-HiPACC) and were run at the NASA Ames Research Center. The MultiDark-Planck (MDPL) and the BigMD simulation suite have been performed in the Supermuc supercomputer at LRZ using time granted by PRACE. E.D.F. thanks Penn State's Center for Astrostatistics for an environment where cross-disciplinary research can be effectively pursued.

\bibliography{Mixture_models_accepted}
\bibliographystyle{aasjournal}

\end{document}